\documentclass{aa}

\usepackage{graphicx}

\usepackage[varg]{txfonts}
\usepackage{natbib}
\bibpunct{(}{)}{;}{a}{}{,}
\usepackage{array}
\usepackage{xspace}
\usepackage{lscape}
\usepackage{url}
\usepackage{arydshln}
\usepackage[hyperfootnotes=false, linktocpage=true, breaklinks=true, colorlinks=true, linkcolor=blue, citecolor=blue, urlcolor=blue]{hyperref}
\usepackage[all]{hypcap}
\usepackage{longtable}
\usepackage{afterpage}

\usepackage[dvipsnames]{xcolor}

\newcommand{\FeKa}{Fe K\ensuremath{\alpha}\xspace}

\newcommand{\kms}{\ensuremath{\mathrm{km\ s^{-1}}}\xspace}
\newcommand{\NH}{\ensuremath{N_{\mathrm{H}}}\xspace}

\newcommand{\xmm}{{\it XMM-Newton}\xspace}

\newcommand{\spitzer}{{\it Spitzer}\xspace}

\newcommand{\suzaku}{{\it Suzaku}\xspace}

\newcommand{\athena}{{\it Athena}\xspace}

\newcommand{\ergflux}{{\ensuremath{\rm{erg\ cm}^{-2}\ \rm{s}^{-1}}}\xspace}
\newcommand{\ergs}{{\ensuremath{\rm{erg\ s}^{-1}}}\xspace}
\newcommand{\cm}{{\ensuremath{\rm{cm}^{-2}}}\xspace}

\newcommand{\spex}{\xspace{\tt SPEX}\xspace}

\newcommand{\MBH}{\ensuremath{{M_{\rm BH}}}\xspace}

\mathchardef\mhyphen="2D

\begin{document}

\title{Relation between winds and jets in radio-loud AGN}

\author{
Missagh Mehdipour
\and
Elisa Costantini
}
\institute{
SRON Netherlands Institute for Space Research, Sorbonnelaan 2, 3584 CA Utrecht, the Netherlands\\ \email{M.Mehdipour@sron.nl}
}
\date{Received 5 February 2019 / Accepted 26 March 2019}
\abstract
{
We investigate the relation between the two modes of outflow (wind and jet) in radio-loud active galactic nuclei (AGN).
For this study we have carried out a systematic and homogeneous analysis of {\it XMM-Newton} spectra of a sample of 16 suitable radio-loud Seyfert-1 AGN. 
The ionised winds in these AGN are parameterised through high-resolution X-ray spectroscopy and photoionisation modelling.
We discover a significant inverse correlation between the column density $N_{\rm H}$ of the ionised wind and the radio-loudness parameter $R$ of the jet.
We explore different possible explanations for this $N_{\rm H}$-$R$ relation and find that ionisation, inclination, and luminosity effects are unlikely to be responsible for the observed relation.
We argue that the $N_{\rm H}$-$R$ relation is rather a manifestation of the magnetic driving mechanism of the wind from the accretion disk. 
Change in the magnetic field configuration from toroidal to poloidal, powering either the wind or the jet mode of the outflow, is the most feasible explanation for the observed decline in the wind \NH as the radio jet becomes stronger.
Our findings provide evidence for a wind-jet bimodality in radio-loud AGN and shine new light on the link between these two modes of outflow. This has far-reaching consequences for the accretion disk structure and the wind ejection mechanism.
}
\authorrunning{M. Mehdipour \& E. Costantini}
\titlerunning{Relation between winds and jets in radio-loud AGN}
\maketitle
\section{Introduction}

The growth of supermassive black holes (SMBHs) via accretion is linked to the growth of their host galaxy via star formation. This is observed as a correlation between the SMBH mass and the stellar mass of the galactic bulge (the so-called M-$\sigma$ relation, e.g. \citealt{Ferr00}). This suggests that SMBHs and their host galaxies are co-evolved. However, the mechanism required for this co-evolution is rather uncertain. In AGN, accretion onto the SMBH liberates enough power to affect the host galaxy, but how this is transported from the very small scales close to the SMBH to the much larger scales of the host galaxy is poorly understood.

Apart from the radiation, accretion is ubiquitously accompanied by outflows (jets and/or winds), which transport matter and energy away from the nucleus into the galaxy (see e.g. the recent reviews by \citealt{King15,Morg17}). These outflows can provide an efficient coupling between SMBHs and their host galaxies. They can impact star formation, chemical enrichment of the surrounding intergalactic medium, and the cooling flows at the core of galaxy clusters (e.g. \citealt{Fabi12}). AGN outflows can therefore play an important role in the co-evolution of SMBHs and their host galaxies through a feedback mechanism. However, there are significant gaps in our understanding of the outflow phenomenon in AGN, which cause uncertainties in determining their role and impact in galaxy evolution.

Observations show that about 15--20\% of AGN are radio-loud \citep{Kell89,Urry95}, where the ratio of radio to optical emission is orders of magnitude higher than in typical radio-quiet AGN. In radio-loud AGN accretion onto the SMBH involves energy release in the form of radio synchrotron radiation from collimated jets of relativistic particles from the central engine. The mechanism for the production of the jet has been thought to be electromagnetic extraction of rotational energy from a spinning black hole via magnetic field lines that are supported by the accretion disk \citep{Blan77,Blan82}. However, the connection between the accretion disk and the jet, and the formation of the observed dichotomy of radio-quiet and radio-loud AGN, are still open questions. Apart from the collimated jets, another important mode/component of AGN outflows is the winds of ionised gas. Importantly, the relationship of the ionised winds to the jet and the disk is not known, which we investigate in this paper.

While jets are generally understood to be magnetically powered, the origin and launching mechanism of the ionised winds in AGN are more uncertain. The winds may originate from either the accretion disk or the torus of the AGN, with different possible mechanisms postulated for their ejection. They could be either thermally-driven (e.g. \citealt{Bege83,Krol01}), or radiatively-driven (e.g. \citealt{Prog04,Higg14}), or magnetically-driven (e.g. \citealt{Koni94,Fuku10}) winds. However, it is still debatable what physical conditions and factors govern the launch of outflows, and what is the dependence on the accretion flows, radiative power, and the mass and the spin of SMBHs. 

The X-ray energy band is where the ionised winds mainly reveal their diagnostic-rich spectral features. The nearby X-ray bright radio-quiet AGN have served as excellent laboratories to study these winds (e.g. \citealt{Kaa00,Kaas12,Blu05,Laha14,Beh17,Meh18b}). However, there are also detections of ionised winds in radio-loud AGN (e.g. \citealt{Reyn97,Ball05b,Reev09b,Torr12,Tomb14,DiGe16}).

In an X-ray study of a sample of Type-1 AGN using Advanced Satellite for Cosmology and Astrophysics (ASCA, \citealt{Tana94}), \citet{Reyn97} suggested that the four radio-loud AGN in their sample had relatively weaker absorption compared to the radio-quiet AGN. However, the number of radio-loud AGN, and the spectral resolution of ASCA, were too limited for a conclusive interpretation. More recently, in a search for ultra-fast outflows at hard X-rays using \xmm \citep{Jans01} and \suzaku \citep{Mits07}, \citet{Tomb14} find evidence of Fe K absorption with relativistic speeds in some radio-loud AGN. They report that the ultra-fast outflows are detected at a wide range of jet inclination angles (${\sim 10\degr}$ to 70$\degr$), suggesting that the winds have a large opening angle and are not preferentially equatorial.

In this paper we investigate the link between the parameters of the wind and the jet in a sample of Type-1 AGN that are X-ray bright and radio loud. We systematically model all the X-ray absorption by the ionised winds in these objects using \xmm (RGS and EPIC spectroscopy), and derive the wind parameters from our homogeneous photoionisation modelling. This is the first time changes in the ionised AGN wind as a function of the jet radio power are investigated. Our characterisation of the wind in radio-loud AGN provides new information for testing theoretical models of the wind-jet-disk connection in AGN.

The structure of the paper is as follows. In Sect. \ref{data_sect} we describe the selection criteria of our AGN sample, how the properties of our AGN sample were collated, as well as the reduction and processing of the \xmm data. The fitting of the X-ray spectra and photoionisation modelling of the AGN wind are reported in Sect. \ref{model_sect}. The wind-jet correlation analysis from the results of our modelling are reported in Sect. \ref{analysis_sect}. We discuss and interpret our findings in Sect. \ref{discussion}, and give concluding remarks in Sect. \ref{conclusions}.

The spectral analysis and modelling presented in this paper were done using the {\tt SPEX} package \citep{Kaa96,Kaas17} v3.04.00 and its {\tt pion} photoionisation model \citep{Meh16b}. We use C-statistics for spectral fitting with the X-ray spectra optimally binned according to \citet{Kaas16}. Errors on the derived parameters are reported at the $1\sigma$ confidence level. We assume proto-solar abundances of \citet{Lod09} throughout our modelling in this paper. In our computations we adopt the cosmological parameters ${H_{0}=70\ \mathrm{km\ s^{-1}\ Mpc^{-1}}}$, $\Omega_{\Lambda}=0.70$, and $\Omega_{\rm m}=0.30$.

\section{Observational data}
\label{data_sect}
%

\subsection{Sample selection}
\label{sample_sect}

We started with the Quasars and Active Galactic Nuclei Catalogue (13th Ed.) of \citet{Vero10} for selecting our AGN and cross matching with \xmm observations. This catalogue provides a collection of useful information on a large number of known AGN (about 169 thousands), such as the spectral classification (e.g. Seyfert 1, BL Lac, etc.), the 6 cm (5 GHz) radio flux density ${F_{\rm 6\, cm}}$, as well as the optical magnitude. From these we calculated the radio-loudness parameter $R$ \citep{Kell89}, which is defined as the ratio of radio to optical flux density: ${R = F_{\rm 6\, cm} / F_{\rm opt}}$. For the calculation of ${F_{\rm opt}}$ in units of erg~cm$^{-2}$~s$^{-1}$~Hz$^{-1}$ from the $B$ magnitude (at 4400 $\AA$) we used the relation ${B = -2.5 \log F_{\rm opt} - 48.36}$ \citep{Schm83} in \citet{Kell89}. The radio-loud AGN are formally classed as those with ${R > 10}$ \citep{Kell89}.

\begin{figure}[!tbp]
\centering
\resizebox{1.01\hsize}{!}{\includegraphics[angle=0]{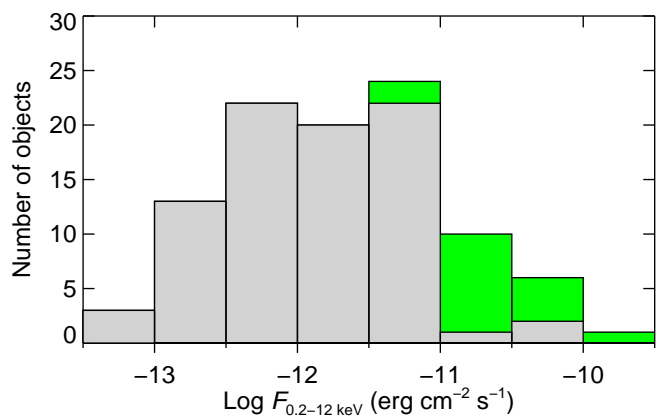}\hspace{-0.0cm}}
\caption{Distribution of the X-ray flux of the radio-loud Type-1 AGN (Seyferts and quasars) obtained from cross-matching the \citet{Vero10} and the \xmm Serendipitous Source Catalogue. The green parts of the histogram correspond to the 16 AGN that meet the selection criteria for having sufficient S/N in the soft X-ray band for high-resolution spectroscopy of the wind with \xmm RGS. Details of the selection criteria are given in Sect. \ref{sample_sect}. Properties of the 16 selected AGN are listed in Table \ref{properties_table}.}
\label{histo_fig}
\end{figure}

We cross matched the AGN in \citet{Vero10} catalogue with the \xmm Serendipitous Source Catalogue 3XMM-DR6 (XMM-SSC, 2016), in order to identify the associated X-ray emission of the AGN in the \citet{Vero10} catalogue, and acquire their X-ray fluxes. For the cross matching of the two catalogues we used the CDS X-Match Service \citep{Pine11} provided by the Centre de Donn\'ees astronomiques de Strasbourg (CDS). The cross matching of the two catalogues resulted in about 8800 matched X-ray sources. Following this, we proceeded to filter out those objects classed by \citet{Vero10} as BL-Lacs, highly polarised objects, LINERs, and Type-2 AGN. This left us with about 2400 Type-1 AGN remaining, which consist of Seyfert-1 galaxies and quasars. We then proceeded to select only the radio-loud AGN. Thus, for the objects with available ${F_{\rm 6\, cm}}$ and $B$ data, those with ${R > 10}$ were selected. This gave us 99 radio-loud Type-1 AGN. The distribution of the X-ray flux of these AGN is shown in the histogram of Fig. \ref{histo_fig}. In order to model the AGN wind and derive its parameters, high-resolution X-ray spectroscopy of the absorption features in the soft X-ray band is needed. Therefore, RGS spectrum with sufficient signal-to-noise ratio (S/N) is required. We thus selected only those AGN with S/N of at least 3 in the RGS resolution element (about 70 m$\AA$) at the continuum level. This means those AGN with extreme neutral absorption (either Galactic or in the host galaxy), or intrinsically faint X-ray continuum, or those with insufficient \xmm exposure are effectively filtered out. The 16 remaining AGN that are suitable for the wind study are highlighted in green in the histogram of Fig. \ref{histo_fig}. The properties of these selected radio-loud Type-1 AGN are listed in Table \ref{properties_table}.

%

%
\begin{table*}[!t]
\begin{minipage}[t]{\hsize}
\setlength{\extrarowheight}{3pt}
\caption{Properties of the 16 radio-loud AGN in our sample.}
\label{properties_table}
\footnotesize
\renewcommand{\footnoterule}{}
\centering
\begin{tabular}{c c c c c c c c c c c c c c c c}
\hline \hline
         Object  &       Class  &         $z$  &         $B$  &      $F_{\rm mid \mhyphen IR}$  &      $F_{6\, \rm cm}$  &             $R$  &         $i$  &      $\MBH$  &              Gal \NH  &        $F_{\rm soft}$  &        $F_{\rm hard}$  &   $L_{\rm X}$   &   $L_{\rm bol}$  & $\lambda$   \\
            (1)  &         (2)\tablefootmark{a}  &         (3)\tablefootmark{a}  &         (4)\tablefootmark{a}  &                   (5)  &                   (6)  &                   (7)\tablefootmark{k}  &         (8)  &         (9)  &                  (10)\tablefootmark{ah}  &                  (11)\tablefootmark{k}  &                  (12)\tablefootmark{k}  &                  (13)\tablefootmark{k}  &     (14)\tablefootmark{k} &   (15)\tablefootmark{k}   \\
         \hline
      \object{1H 0323+342}  &         S1n  &       0.063  &       15.72  &                  n/a  &                  0.30\tablefootmark{d}  &                  130  &         13\tablefootmark{l}  &       0.36\tablefootmark{v}  &                  21.7  &                  0.53  &                  0.81  &       0.278          & 2.08  &                      0.46   \\         
          \object{3C 59}  &        S1.8  &       0.111  &       16.80  &                  n/a  &                  0.02\tablefootmark{d}  &                   22  &         n/a  &        7.94\tablefootmark{w}  &                  6.59  &                  0.36  &                  0.86  &      0.538        &    3.88  &                         0.04   \\
         \object{3C 120}  &        S1.5  &       0.033  &       15.72  &                  0.54\tablefootmark{b}  &                  0.36\tablefootmark{e}  &                  154  &         21\tablefootmark{m}  &        0.69\tablefootmark{x}  &                  19.4  &                  3.03  &                  5.27  &     0.309    &         2.33  &                          0.27   \\
         \object{3C 273}  &        S1.0  &       0.158  &       13.05  &                  0.60\tablefootmark{b}  &                 38.4\tablefootmark{f}  &                 1407  &         20\tablefootmark{n}  &       65.9\tablefootmark{y}  &                  1.77  &                  5.34  &                  7.93  &     9.16     &        97.4  &                          0.12   \\
         \object{3C 382}  &        S1.0  &       0.058  &       16.50  &                  0.09\tablefootmark{b}  &                  0.15\tablefootmark{d}  &                  128  &         40\tablefootmark{o}  &       11.5\tablefootmark{z}  &                  8.96  &                  2.03  &                  3.44  &     0.544      &       4.07  &                         0.03   \\
       \object{3C 390.3}  &        S1.5  &       0.056  &       16.06  &                  0.22\tablefootmark{b}  &                  0.12\tablefootmark{e}  &                   70  &         33\tablefootmark{o}  &        2.87\tablefootmark{aa}  &                  4.41  &                  2.10  &                  3.48  &     0.459       &      3.36  &                         0.09   \\
      \object{4C +31.63}  &        S1.0  &       0.298  &       15.85  &                  0.07\tablefootmark{b}  &                  1.40\tablefootmark{d}  &                  676  &         10\tablefootmark{p}  &       20.0\tablefootmark{ab}  &                  12.0  &                  0.28  &                  0.45  &     3.28        &     29.0  &                          0.11   \\ 
      \object{4C +34.47}  &        S1.0  &       0.206  &       15.58  &                  0.05\tablefootmark{b}  &                  0.37\tablefootmark{d}  &                  139  &         35\tablefootmark{q}  &        3.16\tablefootmark{ac}  &                  3.36  &                  0.70  &                  0.86  &     2.19         &    16.4  &                          0.41   \\
      \object{4C +74.26}  &        S1.0  &       0.104  &       15.13  &                  0.14\tablefootmark{b}  &                  0.32\tablefootmark{g}  &                   79  &         40\tablefootmark{r}  &       41.7\tablefootmark{ad}  &                  23.1  &                  0.85  &                  2.46  &     1.41          &   10.4  &                         0.02   \\
    \object{ESO 075-G041}  &          S1  &       0.028  &       14.78  &                  0.12\tablefootmark{c}  &                 13.4\tablefootmark{h}  &                 2416  &         10\tablefootmark{s}  &       n/a  &                  2.90  &                  0.49  &                  0.67  &          0.023     &  0.31  &                      n/a   \\      
       \object{III Zw 2}  &        S1.2  &       0.089  &       15.96  &                  0.13\tablefootmark{b}  &                  0.43\tablefootmark{d}  &                  233  &         21\tablefootmark{q}  &        0.72\tablefootmark{x}  &                  7.13  &                  0.39  &                  0.72  &      0.251      &      1.94  &                          0.21   \\
          \object{Mrk 6}  &        S1.5  &       0.019  &       15.16  &                  0.55\tablefootmark{b}  &                  0.10\tablefootmark{d}  &                   26  &         9\tablefootmark{t}  &        1.80\tablefootmark{ae}  &                  9.80  &                  0.17  &                  1.51  &      0.026       &    0.23  &                         0.01   \\
        \object{Mrk 896}  &         S1n  &       0.027  &       15.27  &                  0.13\tablefootmark{c}  &                  0.04\tablefootmark{i}  &                   11  &         24\tablefootmark{u}  &        0.12\tablefootmark{af}  &                  4.03  &                  0.54  &                  0.37  &      0.021        &   0.15  &                          0.10   \\
    \object{PKS 0405-12}  &        S1.2  &       0.574  &       15.09  &                  0.09\tablefootmark{b}  &                  1.99\tablefootmark{j}  &                  477  &         n/a  &       29.5\tablefootmark{ad}  &                  4.16  &                  0.39  &                  0.43  &            13.4      & 112  &                      0.30   \\
   \object{PKS 0921-213}  &          S1  &       0.053  &       16.50  &                  n/a  &                  0.42\tablefootmark{h}  &                  369  &         n/a  &        0.79\tablefootmark{w}  &                  5.75  &                  0.41  &                  0.68  &            0.083    & 0.67  &                         0.07   \\
    \object{PKS 2135-14}  &        S1.5  &       0.200  &       15.63  &                  0.11\tablefootmark{b}  &                 1.33\tablefootmark{h}  &                 525  &         n/a  &     44.7\tablefootmark{ag}  &                5.22  &                 0.36  &                  0.60  &                  1.21  &                 10.2  &                       0.02   \\
\hline
\end{tabular}
\end{minipage}
\tablefoot{ 
(1) Name of the AGN. (2) Spectral classification. (3) Cosmological redshift. (4) $B$ magnitude. (5) Mid-IR (MIPS/24 $\mu$m or IRAS/25$\mu$m) flux in Jy. (6) 6 cm (5 GHz) flux in Jy. (7) Radio-loudness parameter $R$. (8) Inclination angle of the radio jet axis relative to our line of sight in degrees. (9) Mass of the SMBH in $10^{8} M_{\odot}$. (10) Milky Way neutral \NH in our line of sight in $10^{20}$ \cm. (11) Soft X-ray (0.3--2 keV) observed flux in $10^{-11}$~\ergflux derived from our \xmm spectral modelling. (12) Hard X-ray (2--10 keV) observed flux in $10^{-11}$~\ergflux derived from our \xmm spectral modelling. (13) Intrinsic X-ray luminosity over 0.3--10 keV in $10^{45}$~\ergs derived from our \xmm spectral modelling. (14) Intrinsic bolometric luminosity of the AGN in $10^{45}$~\ergs. (15) The Eddington ratio ${\lambda = {L_{\rm bol} / L_{\rm Edd}}}$. References: 
\tablefoottext{a}{\citet{Vero10}.}
\tablefoottext{b}{\spitzer MIPS/24$\mu$m from IRSA.}
\tablefoottext{c}{IRAS/25$\mu$m from \citet{Mosh90}.}
\tablefoottext{d}{\citet{Laur97}.}
\tablefoottext{e}{\citet{Dods08}.}
\tablefoottext{f}{\citet{Gear94}.}
\tablefoottext{g}{\citet{Kueh81}.}
\tablefoottext{h}{\citet{Wrig90}.}
\tablefoottext{i}{\citet{Bica95}.}
\tablefoottext{j}{\citet{Shim72}.}
\tablefoottext{k}{Calculated in this paper.}
\tablefoottext{l}{\citet{Karam15}.}
\tablefoottext{m}{\citet{Jors05}.}
\tablefoottext{n}{\citet{Staw04}.}
\tablefoottext{o}{\citet{Giov01}.}
\tablefoottext{p}{\citet{Hoga11}.}
\tablefoottext{q}{\citet{Roka03}.}
\tablefoottext{r}{\citet{Ball05a}.}
\tablefoottext{s}{\citet{Worr12}.}
\tablefoottext{t}{\citet{Khar06}.}
\tablefoottext{u}{\citet{Page03}.}
\tablefoottext{v}{\citet{Fosc15}.}
\tablefoottext{w}{\citet{Siko07}.}
\tablefoottext{x}{\citet{Grie17}.}
\tablefoottext{y}{\citet{Palt05}.}
\tablefoottext{z}{\citet{Marc04}.}
\tablefoottext{aa}{\citet{Pet04}.}
\tablefoottext{ab}{\citet{Veil09}.}
\tablefoottext{ac}{\citet{Vasu07}.}
\tablefoottext{ad}{\citet{Woo02}.}
\tablefoottext{ae}{\citet{Doro12}.}
\tablefoottext{af}{\citet{Piot15}.}
\tablefoottext{ag}{\citet{Ming14}.}
\tablefoottext{ah}{\citet{Will13}.}
}
\end{table*}

\subsection{Properties of the selected AGN}
\label{parameter_sect}

Following the selection of our AGN sample in Sect. \ref{sample_sect}, we compiled information on properties of these AGN from the literature. These properties which are listed in Table \ref{properties_table}, alongside the results of our modelling of the AGN wind (Sect. \ref{model_sect}), are required for our investigation of the wind-jet-disk connection. We describe below how the properties in Table \ref{properties_table} were obtained.

The classification, the cosmological redshift $z$, and the $B$ magnitude of all the AGN in our sample are taken from the \citet{Vero10} catalogue and references therein. The radio and X-ray fluxes were initially taken from \citet{Vero10} and the \xmm Serendipitous Source Catalogue, respectively, for the purpose of selecting the AGN (Sect. \ref{sample_sect}). However, these flux measurements were further refined for our study in this paper, which are provided in Table \ref{properties_table} with their references. The soft (0.3--2 keV) and hard (2--10 keV) observed X-ray fluxes, and the intrinsic X-ray luminosity in Table \ref{properties_table}, are derived from our spectral modelling presented in Sect. \ref{model_sect}. 

In order to assess the strength of emission from the AGN dusty torus we obtained the mid-IR flux of the objects in our sample. We used mid-IR flux measurements from \spitzer Multi-band Imaging Photometer (MIPS) taken in the 24 $\mu$m photometric channel. The MIPS fluxes were retrieved from the Spitzer Enhanced Imaging Products (SEIP) source list through NASA/IPAC Infrared Science Archive (IRSA). For two of the AGN in our sample (ESO~075-G041 and Mrk~896), which \spitzer/MIPS observations were not available, we used mid-IR flux measurements from Infrared Astronomical Satellite (IRAS) taken in the 25 $\mu$m photometric channel. The IRAS fluxes were obtained from the IRAS Faint Source catalogue v2.0 \citep{Mosh90}. The mid-IR flux ($F_{\rm mid \mhyphen IR}$) of the AGN in our sample are provided in Table \ref{properties_table}. 

We also collected from the literature the inclination angle $i$ of the radio jet axis relative to our line of sight, and the mass of the SMBH ($M_{\rm BH}$) in the AGN. For one of the objects in our sample (Mrk~896), where the jet inclination angle is not available, we instead take the reported inclination angle of the accretion disk. The $i$ and $M_{\rm BH}$ values of our AGN sample, and their reference papers, are provided in Table \ref{properties_table}. The intrinsic bolometric luminosity $L_{\rm bol}$ of the AGN was estimated using a spectral energy distribution (SED) model. As the objects in our sample are radio-loud Seyfert-1 AGN, we use a radio-loud version of the SED model derived from a multi-wavelength campaign study of the archetypal Seyfert-1 AGN \object{NGC 5548} \citep{Meh15a}. We modified this radio-quiet SED, which spans from near-IR to hard X-rays, by extending it to lower energies using a power-law model for the synchrotron emission from the jet. In our model this power-law continuum spans from near-IR (1 $\mu$m) to radio (100 MHz) energies with a fiducial photon index of 1.8. For radio-loudness parameter $R$ ranging from 10 to $10^{3}$, the corresponding luminosity contribution by the power-law extension ranges from $10^{42}$ to $10^{44}$ \ergs. This radio-loud broadband SED model was used to estimate the bolometric luminosity $L_{\rm bol}$ by matching the intrinsic X-ray luminosity $L_{\rm X}$ and the radio flux $F_{\rm 6\, cm}$ of each AGN in our sample. Finally, from $M_{\rm BH}$ and $L_{\rm bol}$, the Eddington ratio $\lambda$ was calculated. The $L_{\rm X}$, $L_{\rm bol}$, and the $\lambda$ values of the AGN in our sample are shown in Table \ref{properties_table}.

\subsection{X-ray data reduction and processing}

The \xmm Observation Data Files (ODFs) of the AGN in our sample were retrieved from the \xmm Science Archive (XSA) and processed using the Science Analysis System (SAS v17.0.0). The ID and the duration of the \xmm observations that we used in our study are provided in Table \ref{continuum_table}. For those AGN with multiple \xmm observations we have used one with sufficient S/N for line spectroscopy of the wind with RGS. The data from the RGS instruments \citep{denH01} were processed through the {\tt rgsproc} pipeline task; the source and background spectra were extracted and the response matrices were generated. We filtered out time intervals with background count rates $> 0.1\ \mathrm{count\ s}^{-1}$ in CCD number 9. The {\tt rgscombine} task was used to stack the first-order RGS1 and RGS2 spectra of each \xmm observation. The fitted spectral range is 7--36~\AA\ for RGS. The EPIC-pn \citep{Stru01} and EPIC-MOS \citep{Turn01} data were processed with the {\tt epproc} and {\tt emproc} pipeline tasks, respectively. Periods of high flaring background for EPIC-pn (${ > 0.40}$ count~s$^{-1}$) and EPIC-MOS (${ > 0.35}$ count~s$^{-1}$) were filtered out by applying the {\tt \#XMMEA\_EP} and {\tt \#XMMEA\_EM} filters, respectively. The EPIC spectra were extracted from a circular region centred on the source with a radius of $40''$. The background was extracted from a nearby source-free region of the same size. The pileup was evaluated to be negligible for majority of the AGN in our sample. However, the largest pileup for the brightest object (3C 273) was found to be about 8\% in MOS and 4\% in pn. Nonetheless, for our purpose of the line spectroscopy of the wind in the soft band pileup is not an issue. The single and double events were selected for the pn ({\tt PATTERN <= 4}) and MOS ({\tt PATTERN <= 12}). Response matrices were generated for the EPIC spectra of each observation using the {\tt rmfgen} and {\tt arfgen} tasks. The {\tt epicspeccombine} task was used for stacking the MOS1 and MOS2 spectra of each observation. Our fitted range for the EPIC spectra is from 0.3 to 10 keV. In our spectral modelling (Sect. \ref{model_sect}) in \spex, we simultaneously fit the RGS, EPIC-pn, and EPIC-MOS spectra of each \xmm observation. Our fitting takes into account differences in their instrumental flux normalisation, which ranges from about few \% to 10\%.

\section{Spectral modelling of the X-ray data}
\label{model_sect}
%

\subsection{X-ray continuum}
\label{continuum_sect}

The continuum model that we applied to fit the X-ray spectra in \spex consists of three components: a power-law component ({\tt pow}), a modified black body component ({\tt mbb}), and a neutral reflection component ({\tt refl}). The {\tt pow} component models the primary broadband continuum, while the {\tt mbb} component models the commonly seen `soft X-ray excess', which is an excess emission at $\le 2$~keV in the X-ray spectra of AGN (e.g. \citealt{Meh11}). The {\tt mbb} model in \spex is a blackbody component modified by coherent Compton scattering \citep{Kaa89}. The {\tt refl} model fits the \FeKa line in the EPIC spectra, which is part of the reflection spectrum produced by the reprocessing of the primary X-ray power-law by neutral gas in the AGN. The {\tt refl} model in \spex computes the \FeKa line according to \citet{Zyck94}, and the Compton-reflected continuum according to \citet{Magd95}, as described in \citet{Zyck99}. The normalisation and the photon index $\Gamma$ of the incident power-law for {\tt refl} were set to those of the observed power-law component ({\tt pow}). The exponential high-energy cut-off of {\tt pow} and the incident power-law for {\tt refl} were fixed to a fiducial value of 200~keV. In our modelling we fitted the reflection scale $s$ parameter of the {\tt refl} model. The ionisation parameter of {\tt refl} was set to zero to produce a neutral reflection component, and its abundances were kept at their default solar values. The best-fit parameters of the continuum components ({\tt pow}, {\tt mbb}, and {\tt refl}) for each AGN in our sample are shown in Table \ref{continuum_table}.

\subsection{X-ray absorption by the diffuse ISM}
\label{ISM_sect}

In our spectral modelling we take into account the continuum and line absorption by the diffuse interstellar medium (ISM) in the Milky Way and the host galaxy of the AGN. This is done using the {\tt hot} model in \spex \citep{dePl04,Stee05}. This model calculates the transmission of a plasma in collisional ionisation equilibrium at a given temperature, which for neutral ISM is set to the minimum temperature of the model at 0.5 eV. We applied one {\tt hot} component for the Milky Way and another one for the host galaxy of the AGN. The total Galactic \NH in our line of sight towards each object (provided in Table \ref{properties_table}) was obtained from \citet{Will13} and kept fixed in our modelling. The host galaxy \NH was allowed to be fitted, and its best-fit value (or upper limit) for each object is shown in Table \ref{wind_table}.

\subsection{Photoionisation and X-ray spectral modelling of the ionised wind in AGN}
\label{pion_sect}

The photoionisation modelling of the ionised AGN wind was carried out in \spex using the {\tt pion} model \citep{Meh16b}. We used a template SED for the calculation of the photoionisation equilibrium. The ionising SED of the Seyfert-1 AGN in our sample, extending from optical/UV to hard X-rays, can be reasonably approximated by the SED of the archetypal Seyfert-1 AGN NGC 5548, which was derived in \citet{Meh15a} from continuum modelling of simultaneous multi-wavelength data. The results of the {\tt pion} calculations were then inputted into the {\tt xabs} model in \spex, which calculates the absorption spectrum of a slab of photoionised gas, where all ionic column densities are linked in a physically consistent fashion according to the {\tt pion} run. The parameters of the {\tt xabs} model are the ionisation parameter $\xi$, the hydrogen column density $\NH$, the covering fraction $C_f$, and the outflow $v_{\rm out}$ and the turbulent $\sigma_v$ velocities. The ionisation parameter $\xi$ \citep{Tar69,Kro81} is defined as ${\xi \equiv {L}\, /\, {{n_{\rm{H}} r^2 }}}$, where $L$ is the luminosity of the ionising source over 1--1000 Ryd (13.6 eV to 13.6 keV) in $\rm{erg}\ \rm{s}^{-1}$, $n_{\rm{H}}$ the hydrogen density in $\rm{cm}^{-3}$, and $r$ the distance between the ionised gas and the ionising source (in cm). The covering fraction $C_f$ was fixed to unity in our modelling.

We started with one {\tt xabs} component and fitted its \NH, $\xi$, and $v_{\rm out}$ parameters. This resulted in significant improvement to the spectral fit of all AGN in our sample. For some of the AGN we found that after fitting the first {\tt xabs} component, there were still remaining absorption features at either different ionisation states or outflow velocities that were not fitted. Therefore, for those cases we fitted a second {\tt xabs} component. Two {\tt xabs} components were found to be sufficient to fit all the absorption features, and a third component did not improve the fit and hence was not needed. We kept $\sigma_v$ of the {\tt xabs} components at the default 100~\kms in our modelling in order to limit the number of unnecessary free parameters and also because fitting $\sigma_v$ did not significantly improve the fits. The best-fit parameters of the {\tt xabs} model for the ionised wind in our AGN sample are reported in Table \ref{wind_table}.

%
\begin{table*}[!tbp]
\begin{minipage}[t]{\hsize}
\setlength{\extrarowheight}{3pt}
\caption{Best-fit parameters of the X-ray continuum components in our radio-loud AGN sample, derived from our modelling of the \xmm observations.}
\label{continuum_table}
\footnotesize
\renewcommand{\footnoterule}{}
\centering
\begin{tabular}{c c c c c c c c}
\hline \hline
         Object  &          {\tt pow} Norm.                                   &       {\tt pow} $\Gamma$                                   &          {\tt mbb} Norm.                                   &            {\tt mbb} $T$                                   &           {\tt refl} $s$                                   &               Obs. ID & Obs. t.   \\
            (1)  &                      (2)                                   &                      (3)                                   &                      (4)                                   &                      (5)                                   &                      (6)                                   &                   (7) & (8)   \\
         \hline
    1H 0323+342  & ${                  2.63     \pm                  0.02 }$  & ${                  1.94     \pm                  0.01 }$  & ${                  4.6     \pm                   0.1 }$  & ${                  0.17     \pm                  0.01 }$  & ${                  0.66     \pm                  0.06 }$  &            0764670101 & 81  \\         
          3C 59  & ${                  6.56     \pm                  0.09 }$  & ${                  1.71     \pm                  0.01 }$  & ${                  5.3     \pm                   0.4 }$  & ${                  0.14     \pm                  0.01 }$  & ${                  0.36     \pm                  0.05 }$  &            0205390201 & 82   \\         
         3C 120  & ${                  4.38     \pm                  0.03 }$  & ${                  1.91     \pm                  0.01 }$  & ${                  0.63     \pm                  0.02 }$  & ${                  0.23     \pm                  0.01 }$  & ${                  0.50     \pm                  0.04 }$  &            0693782401 &  29 \\
         3C 273  & ${                113.3     \pm                   0.2 }$  & ${                  1.64     \pm                  0.01 }$  & ${                 37.2     \pm                   0.8 }$  & ${                  0.16     \pm                  0.01 }$  & ${                  0.10     \pm                  0.01 }$  &            0126700801 & 74  \\
         3C 382  & ${                  7.10     \pm                  0.04 }$  & ${                  1.76     \pm                  0.01 }$  & ${                  4.3     \pm                   0.2 }$  & ${                  0.14     \pm                  0.01 }$  & ${                  0.50     \pm                  0.05 }$  &            0790600101 & 31  \\
       3C 390.3  & ${                  5.90     \pm                  0.04 }$  & ${                  1.68     \pm                  0.01 }$  & ${                  0.27     \pm                  0.01 }$  & ${                  0.29     \pm                  0.01 }$  & ${                  0.35     \pm                  0.03 }$  &            0203720301 & 53 \\
      4C +31.63  & ${                 38     \pm                    1 }$  & ${                  1.90     \pm                  0.03 }$  & ${                 30     \pm                    4 }$  & ${                  0.18     \pm                  0.01 }$  & ${                  0.83     \pm                  0.19 }$  &            0550871001 & 25  \\
      4C +34.47  & ${                 28     \pm                    2 }$  & ${                  1.81     \pm                  0.04 }$  & ${                  4.5     \pm                   0.8 }$  & ${                  0.24     \pm                  0.02 }$  & ${                  0.49     \pm                  0.25 }$  &            0102040101  & 8 \\
      4C +74.26  & ${                 20.0     \pm                   0.2 }$  & ${                  1.82     \pm                  0.01 }$  & ${                  1.5     \pm                   0.2 }$  & ${                  0.25     \pm                  0.02 }$  & ${                  0.43     \pm                  0.04 }$  &            0200910201 & 34 \\
    ESO 075-G041  & ${                  0.31     \pm                 0.01 }$  & ${                  1.74     \pm                  0.02 }$  & ${                  0.02     \pm                 0.01 }$  & ${                  0.29     \pm                  0.01 }$  & ${                  0.04     \pm                  0.06 }$  &            0152670101 & 57   \\      
       III Zw 2  & ${                  3.15     \pm                  0.08 }$  & ${                  1.67     \pm                  0.02 }$  & ${                  0.25     \pm                  0.04 }$  & ${                  0.25     \pm                  0.02 }$  & ${                  0.37     \pm                  0.13 }$  &            0127110201 & 16  \\    
          Mrk 6  & ${                  0.22     \pm                 0.01 }$  & ${                  1.47     \pm                  0.02 }$  & ${                  1.7     \pm                   0.9 }$  & ${                  0.10     \pm                  0.01 }$  & ${                  0.42     \pm                  0.05 }$  &            0144230101 & 59 \\
        Mrk 896  & ${                  0.37     \pm                 0.01 }$  & ${                  2.34     \pm                  0.01 }$  & ${0.6 \pm 0.3}$  & ${0.09 \pm 0.01}$  & ${                  1.48     \pm                  0.10 }$  &            0112600501   & 11 \\                    
    PKS 0405-12  & ${                172     \pm                    2 }$  & ${                  1.90     \pm                  0.01 }$  & ${                 56     \pm                    3 }$  & ${                  0.21     \pm                  0.01 }$  & ${                  0.42     \pm                  0.06 }$  &            0202210301 & 82  \\
   PKS 0921-213  & ${                  1.11     \pm                  0.03 }$  & ${                  1.73     \pm                  0.02 }$  & ${                  0.08     \pm                  0.01 }$  & ${                  0.23     \pm                  0.02 }$  & ${                  0.43     \pm                  0.13 }$  &            0065940501 & 18   \\
    PKS 2135-14  &  ${                 16.2     \pm                   0.2 }$  &  ${                  1.75     \pm                  0.01 }$  & ${                  5.7     \pm                    2.0 }$  &  ${                  0.15     \pm                  0.01 }$  &  ${                  0.58     \pm                  0.10 }$  &           0092850201 &  60   \\   
\hline
\end{tabular}
\end{minipage}
\tablefoot{ 
(1) Name of the AGN. (2) Normalisation of the X-ray power-law component in $10^{52}$ photons s$^{-1}$ keV$^{-1}$ at 1 keV. (3) Photon index of the X-ray power-law component. (4) Normalisation of the modified blackbody component in $10^{34}$ cm$^{1/2}$, which is defined as the emitting area times square root of the electron density. (5) Temperature of the modified blackbody component in keV. (6) Scale parameter of the reflection component. (7) ID of the modelled \xmm observation. (8) Duration of the \xmm observation in ks. The X-ray flux and luminosity corresponding to the above continuum parameters are provided in Table \ref{properties_table} for each AGN.
}
\vspace{0.8cm}
\end{table*}
%
\begin{table*}[!tbp]
\begin{minipage}[t]{\hsize}
\setlength{\extrarowheight}{3pt}
\caption{Best-fit parameters of the ionised AGN wind and the neutral ISM gas in the host galaxy of the radio-loud AGN in our sample, derived from our modelling of the \xmm observations.}
\label{wind_table}
\footnotesize
\renewcommand{\footnoterule}{}
\centering
\begin{tabular}{c c c c c c c c}
\hline \hline
         Object  &                   Wind \NH                                                                                                &            Wind $\log \xi$                                                                                                &                   Wind $v_{\rm out}$                                                                                                &              Neutral \NH                                                                 &     C-stat\,/\,d.o.f.   \\
            (1)  &                      (2)                                                                                                &                      (3)                                                                                                &                      (4)                                                                                                &                      (5)                                                                 &                   (6)   \\
         \hline
    1H 0323+342  & ${                   9     \pm                   2 }$ ,   ${                   7.2     \pm                   0.7 }$  & ${                  2.17     \pm                  0.02 }$ ,  ${                  0.15     \pm                  0.12 }$ &  ${                 -830     \pm                  170 }$ ,  ${                 -880     \pm                  120 }$  & ${  < 0.01}$  &         2296\,/\,1488   \\         
          3C 59  & ${                  40     \pm                   2 }$ ,   ${                  81     \pm                    8 }$  & ${                  1.20     \pm                  0.05 }$ ,  ${                  2.42     \pm                  0.03 }$ &  ${                -3530     \pm                  130 }$ ,  ${                 -1000     \pm                  120 }$  & ${ < 2}$  &         2036\,/\,1477   \\         
         3C 120  & ${                  14     \pm                   3 }$   & ${                  2.65     \pm                  0.04 }$  &  ${                -2160     \pm                  360 }$ & ${  < 0.02}$  &         1949\,/\,1496   \\
         3C 273  & ${                   0.7     \pm                   0.2 }$   & ${                  1.90     \pm                  0.08 }$  &  ${                -3670     \pm                  170 }$  & ${ < 0.01}$  &         2588\,/\,1532   \\
         3C 382  & ${                   7     \pm                   2 }$   & ${                  2.44     \pm                  0.06 }$  &  ${                -1350     \pm                  370 }$  & ${ < 0.02}$  &         1805\,/\,1467   \\
       3C 390.3  & ${                   3.7     \pm                   0.6 }$ ,   ${                  11     \pm                    5}$  & ${                  1.63     \pm                  0.11 }$ ,  ${                  2.77     \pm                  0.06 }$ &  ${                -1550     \pm                  160 }$ ,  ${                   +50     \pm                  100 }$  & ${ < 0.2}$  &         1847\,/\,1481   \\
        4C +31.63  & ${                  11     \pm                   3 }$   & ${                 -0.00     \pm                  0.20 }$  &  ${                 -960     \pm                  200. }$  & ${ < 0.3}$  &         1783\,/\,1423   \\
      4C +34.47  & ${                  31     \pm                  11 }$   & ${                  2.12     \pm                  0.05 }$  &  ${                -1500     \pm                  210 }$  & ${  < 0.3}$  &         1546\,/\,1398   \\ 
      4C +74.26  & ${                  36     \pm                   3 }$ ,   ${                  68     \pm                    8}$  & ${                  1.69     \pm                  0.04 }$ ,  ${                  2.46     \pm                  0.04 }$ &  ${                -1490     \pm                   90 }$ ,  ${                -3000     \pm                   500 }$  & ${ < 0.09}$  &         1986\,/\,1511   \\
    ESO 075-G041  & ${                   7     \pm                   1 }$    & ${                 -0.03     \pm                  0.11 }$  &  ${                 -210     \pm                  180 }$  & ${ < 0.05}$  &         1832\,/\,1415   \\      
       III Zw 2  & ${                   7     \pm                   4 }$    & ${                  2.07     \pm                  0.13 }$  &  ${                -1780     \pm                  670 }$  & ${ < 0.2}$  &         1763\,/\,1468   \\    
          Mrk 6  & ${                 116     \pm                   8 }$    & ${                  1.38     \pm                  0.06 }$ &  ${                -4000     \pm                   500 }$  & ${  27 \pm 4}$  &         2287\,/\,1449   \\
        Mrk 896  & ${                  73     \pm                   6 }$    & ${                  2.23     \pm                  0.02 }$  &  ${                 -130     \pm                  150 }$  & ${ < 0.2}$  &         1679\,/\,1429   \\                    
    PKS 0405-12  & ${                   6     \pm                   2 }$    & ${                  1.71     \pm                  0.17 }$  &  ${                 -130     \pm                  200 }$  & ${ < 0.05}$  &         1715\,/\,1465   \\
   PKS 0921-213  & ${                   8     \pm                   3 }$    & ${                  1.92     \pm                  0.12 }$  &  ${                -3540     \pm                  360 }$ & ${  < 0.2}$  &         1700\,/\,1458   \\
    PKS 2135-14  &  ${  5     \pm                   3 }$  &  ${  2.14     \pm                  0.08 }$ &  ${                -1240     \pm                  530 }$  &  ${  < 0.1 }$  &         1831\,/\,1414    \\   
\hline
\end{tabular}
\end{minipage}
\tablefoot{ 
(1) Name of the AGN. (2) Column density \NH of the ionised wind components ({\tt xabs}) in $10^{20}$ cm$^{-2}$. (3) Logarithm of the ionisation parameter $\xi$ of the ionised wind components ({\tt xabs}) in erg cm s$^{-1}$. (4) Velocity of the outflowing ionised wind components ({\tt xabs}) in \kms. (5) Column density \NH of neutral ISM gas component ({\tt hot}) in the host galaxy of the AGN in $10^{20}$ cm$^{-2}$. (6) Best-fit C-stat over degrees of freedom for the modelled \xmm observation. As the positive and negative uncertainties on the wind parameters from our fits were comparable, we give their average in the table.
}
\end{table*}

%
\begin{figure*}[!tbp]
\centering
\resizebox{0.75\hsize}{!}{\includegraphics[angle=0]{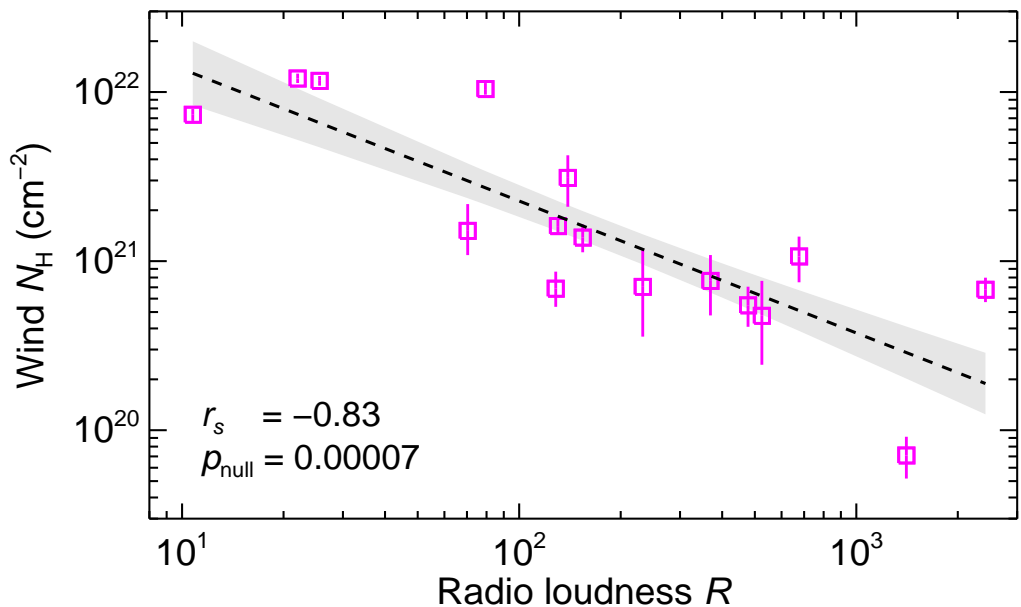}\hspace{-0.0cm}}
\caption{Total column density \NH of the ionised AGN wind versus the radio-loudness parameter $R$ of the jet. Each data point corresponds to one radio-loud AGN in our sample. For those AGN with multiple absorption components the sum of the column density \NH of the components is shown. There is a significant inverse correlation between \NH and $R$. The Spearman's rank correlation coefficient $r_s$ and the null hypothesis probability $p_{\rm null}$ of the correlation between the parameters are shown inset. The data have been fitted with the power-law function of Eq. (\ref{NH_R_eq}) shown in dashed black line. The shaded grey region represents the 1$\sigma$ uncertainty around the best-fit line.}
\label{NH_R_fig}
\end{figure*}

\section{Correlation analysis of the modelling results}
\label{analysis_sect}

Here we examine the results of our modelling of the ionised winds in radio-loud AGN, carried out in Sect. \ref{model_sect}. We look for correlations between the parameters of the wind, jet, disk, and the torus (see Tables \ref{properties_table}, \ref{continuum_table}, and \ref{wind_table}) in order to find out any physical link between them in radio-loud AGN. For our correlation analysis we calculated the Spearman's rank correlation coefficient ($r_{s}$) and the associated null hypothesis p-value probability ($p_{\rm null}$) from the two-sided t-test.

Figure \ref{NH_R_fig} shows the total column density \NH of the ionised wind plotted versus the radio-loudness parameter $R$ of the jet in our AGN sample. Interestingly, it is apparent that \NH decreases as $R$ increases. This inverse correlation is statistically significant with ${r_s = -0.83}$ and probability $p_{\rm null} = 0.00007$. The data in Fig. \ref{NH_R_fig} are fitted with a power-law function, given by
\begin{equation}
\label{NH_R_eq}
{\NH} = 8.2\times 10^{22}\, {{R}^{-0.8}}
\end{equation}
where \NH is the total column density of the ionised wind in \cm, and the radio-loudness parameter of the jet ${R = F_{\rm 6\, cm} / F_{\rm opt}}$.

\begin{figure*}[!tbp]
\centering
\hspace{-0.03cm}\resizebox{1.005\hsize}{!}{\includegraphics[angle=0]{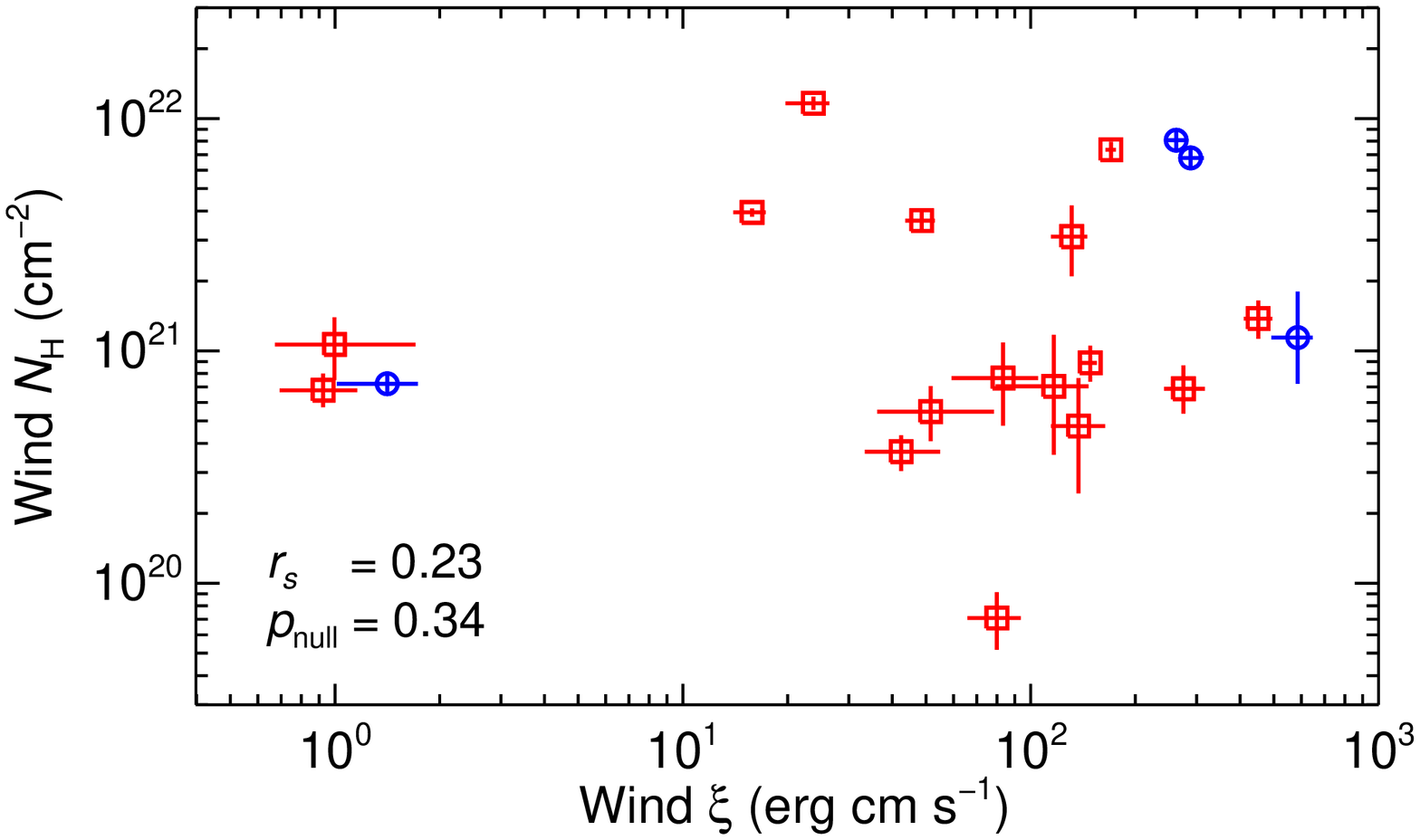}\hspace{-0.0cm} \includegraphics[angle=0]{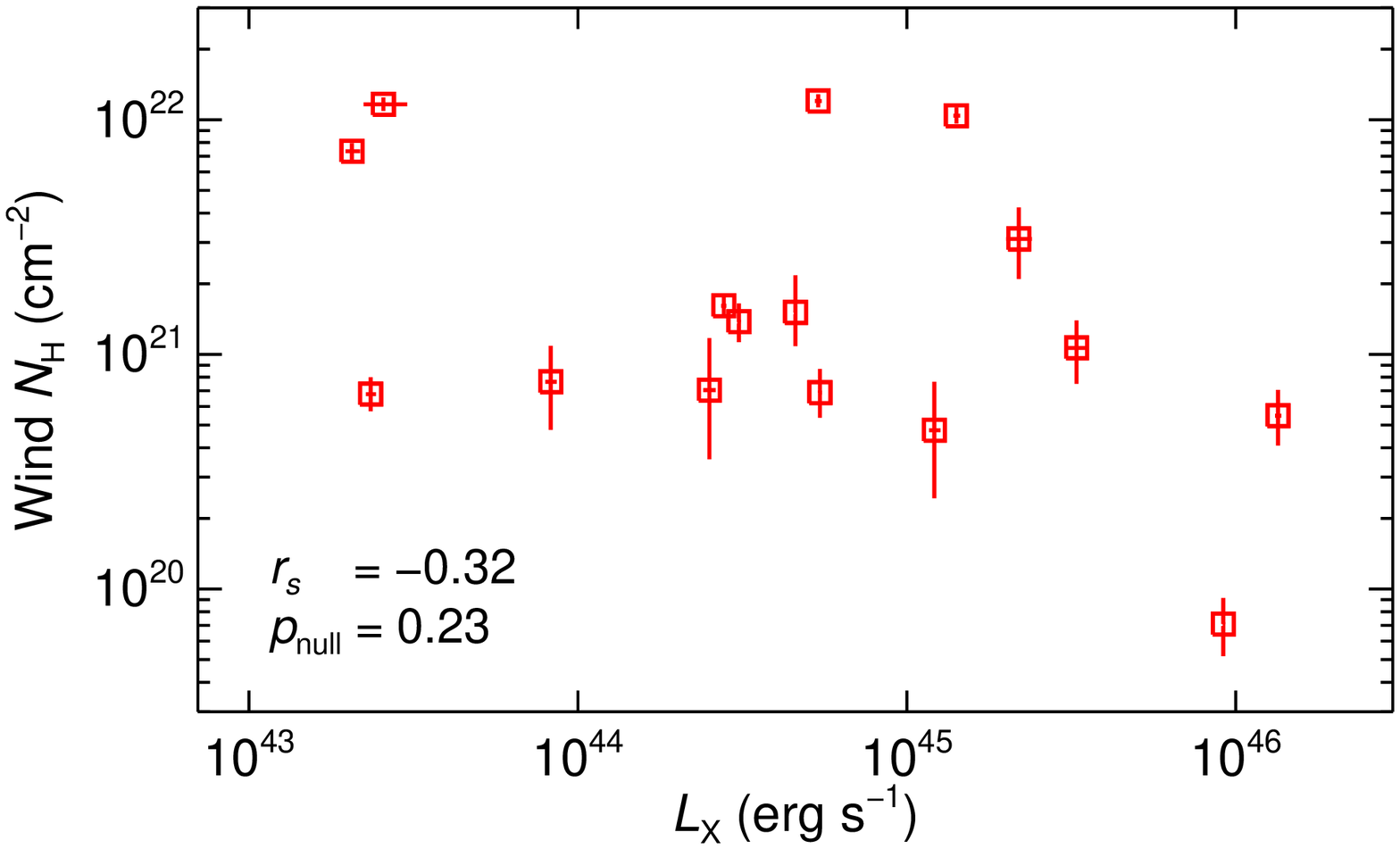}}
\hspace{-0.03cm}\resizebox{1.005\hsize}{!}{\includegraphics[angle=0]{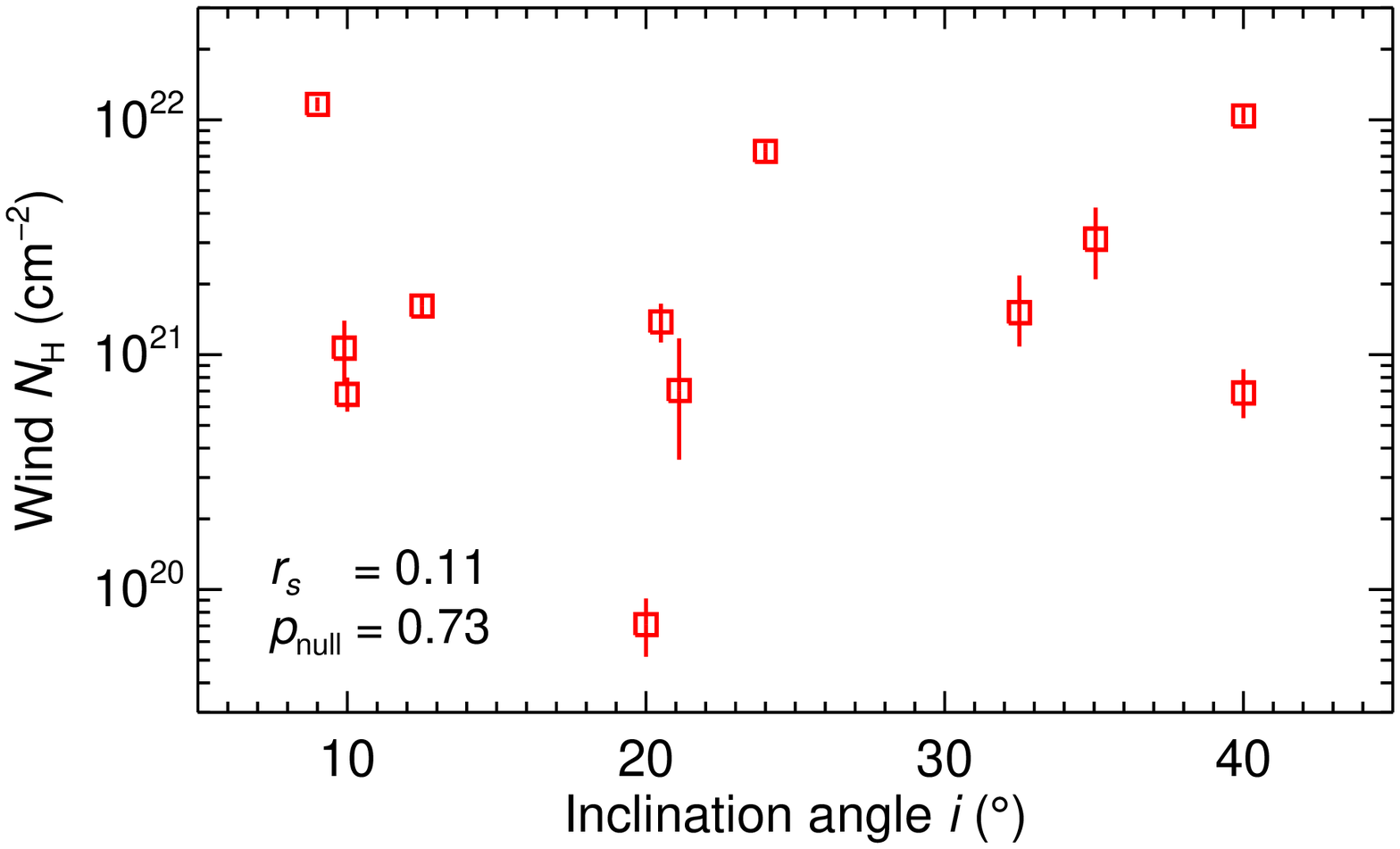}\hspace{-0.0cm} \includegraphics[angle=0]{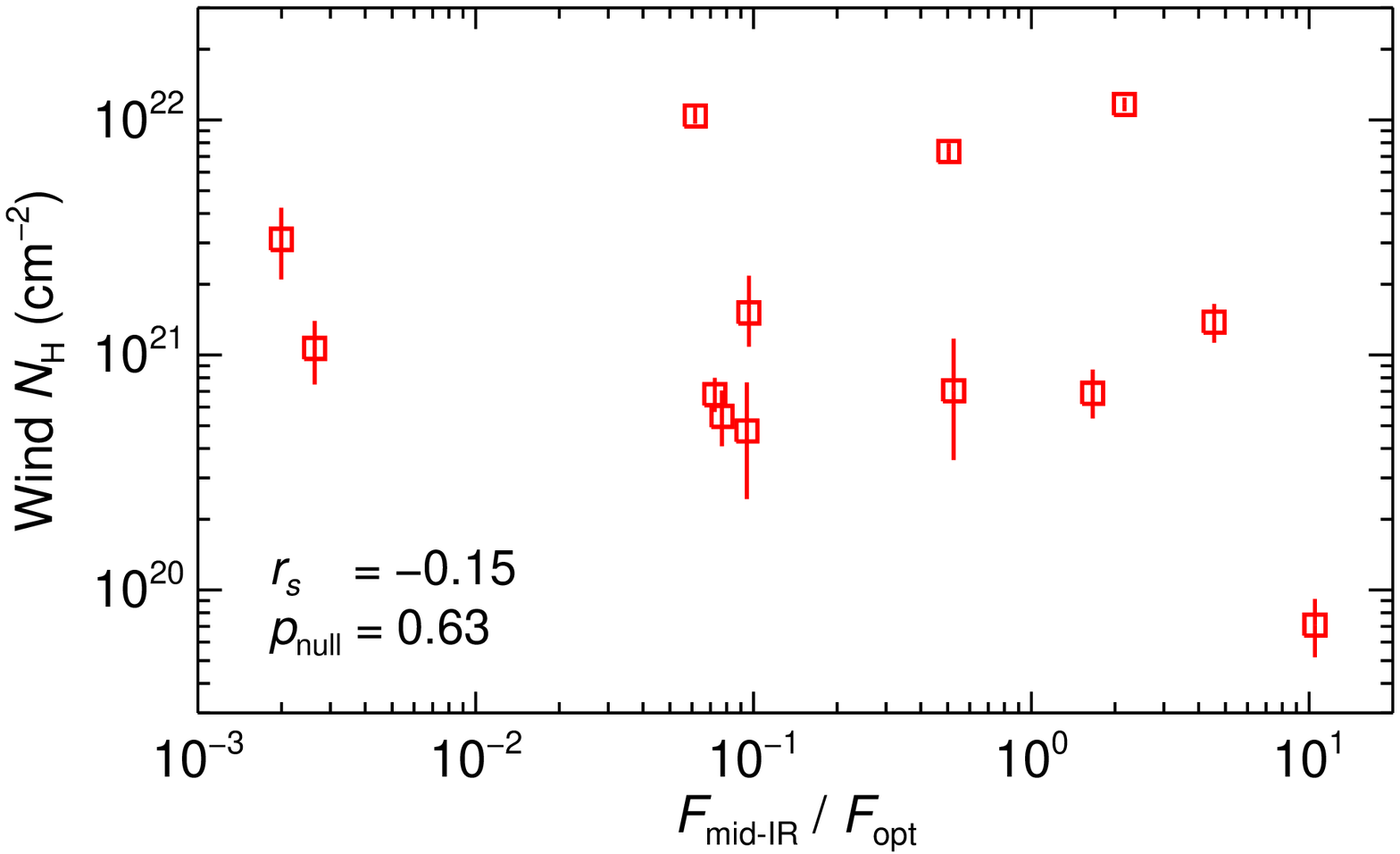}}
\hspace{-0.03cm}\resizebox{1.005\hsize}{!}{\includegraphics[angle=0]{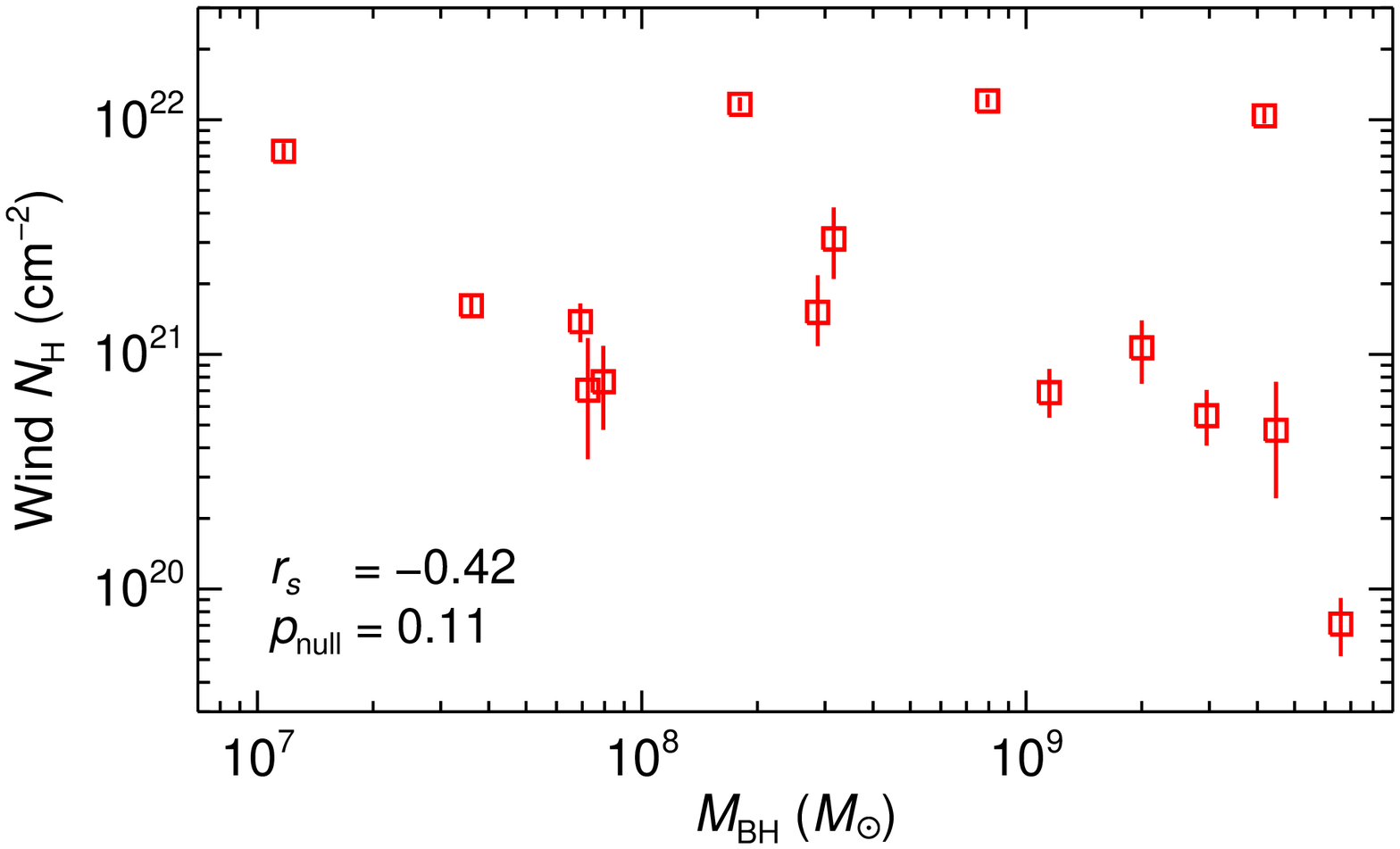}\hspace{-0.0cm} \includegraphics[angle=0]{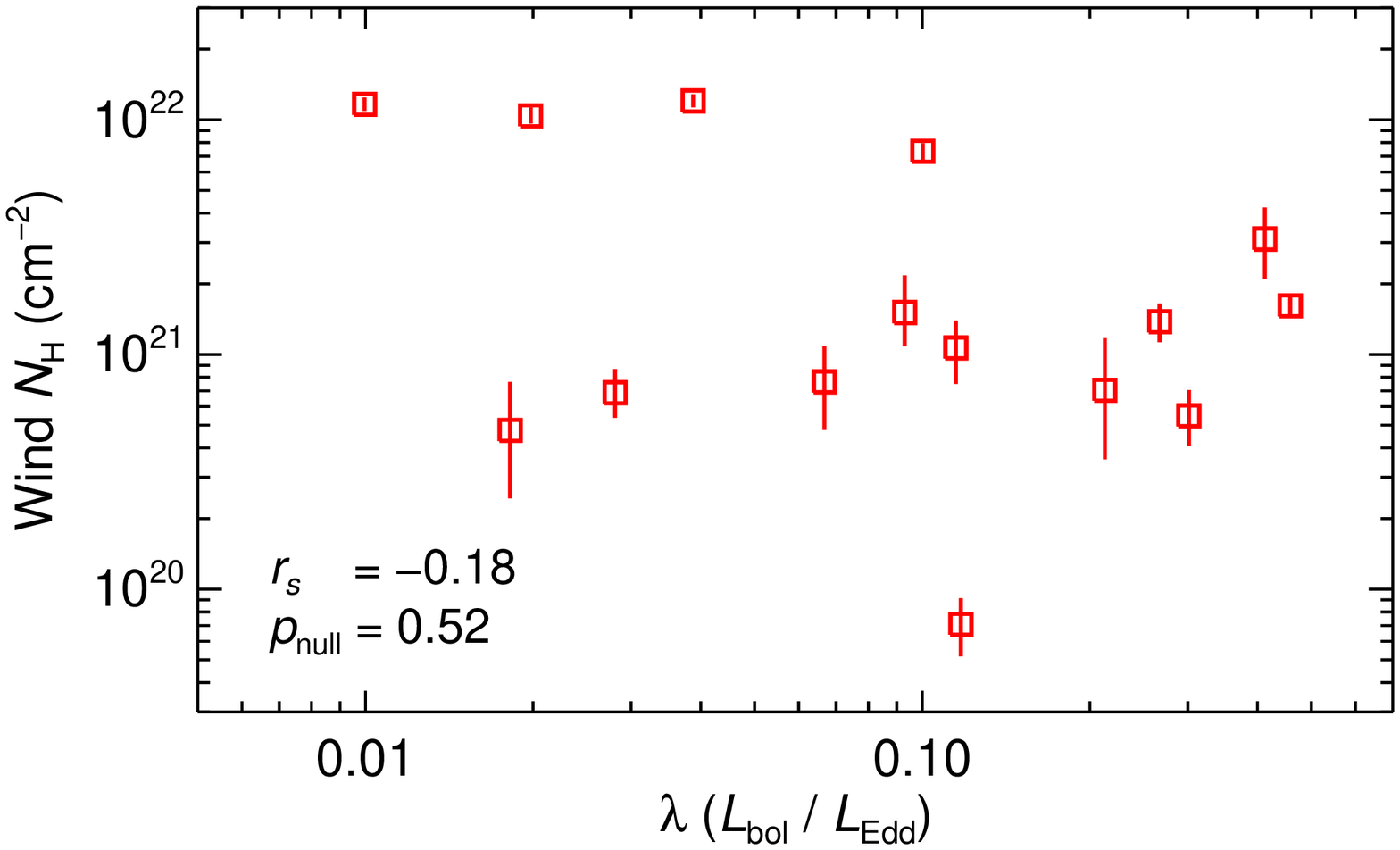}}
\caption{{\it Top left panel}: Column density of the ionised AGN wind components versus their ionisation parameter. The red squares correspond to the first {\tt xabs} component of the wind, while the blue circles correspond to the second {\tt xabs} component for those AGN with two ionised wind components (see Table \ref{wind_table}). {\it Top right panel}: Total column density of the ionised AGN wind versus the intrinsic X-ray luminosity of the AGN over 0.3--10 keV. {\it Middle left panel}: Total column density of the ionised AGN wind versus the inclination angle of the radio jet relative to our line of sight. {\it Middle right panel}: Total column density of the ionised AGN wind versus the ratio of mid-IR to optical flux density (i.e. the `mid-IR loudness' parameter of the dusty torus). {\it Bottom left panel}: Total column density of the ionised AGN wind versus the mass of the SMBH. {\it Bottom right panel}: Total column density of the ionised AGN wind versus the Eddington ratio (ratio of bolometric to Eddington luminosity). The Spearman's rank correlation coefficient $r_s$ and the null hypothesis probability $p_{\rm null}$ of the correlation between the parameters are shown in the corner of each panel.}
\label{NH_rel_fig}
\end{figure*}

There can be different possible explanations for the discovered \NH-$R$ relation, which we will assess and discuss in Sect. \ref{discussion}. In order to aid us identify the physical reason, we investigate the relation between the wind \NH and other key parameters in Fig. \ref{NH_rel_fig}. These parameters are the ionisation parameter $\xi$ of the wind (top left), intrinsic X-ray luminosity of the AGN (top right), the radio jet inclination angle relative to our line of sight (middle left), the ratio of mid-IR to optical flux (middle right), SMBH mass (bottom left), and the Eddington ratio $\lambda$ (bottom right). The $r_{s}$ correlation coefficient and the null hypothesis probability $p_{\rm null}$ are shown in the corner of each panel. The results show no significant correlation between \NH of the wind and the aforementioned parameters. In Sect. \ref{discussion} we discuss the \NH-$R$ relation and the implications of the lack of correlation between \NH and the other parameters.

%

\section{Discussion}
\label{discussion}
%

\subsection{Origin of the observed N\textsubscript{H}-R relation in radio-loud AGN}
\label{explain_sect}

In Sect. \ref{analysis_sect} we found that there is a significant inverse correlation between the wind \NH and the radio-loudness parameter $R$ (Fig. \ref{NH_R_fig}). This shows that as the radio jet becomes stronger, the ionised wind becomes weaker. Importantly, we found no significant correlation between the wind \NH and the other key parameters shown in Fig. \ref{NH_rel_fig}.

The fact that \NH is not related to the ionisation parameter $\xi$ and the X-ray luminosity $L_{\rm X}$, implies that the observed \NH-$R$ relation is not a manifestation of changes in the ionisation state of the wind. This means the lowering of the detected \NH in the soft X-ray band is not caused by a shift in $\xi$ of the gas to higher ionisation states. If the \NH-$R$ relation was a result of changes in $\xi$, then an inverse relation between \NH and $\xi$ would be expected, which is not seen (Fig. \ref{NH_rel_fig}, top left panel). This is also supported by the fact that there is no correlation between \NH and the ionising X-ray luminosity $L_{\rm X}$ of the AGN (Fig. \ref{NH_rel_fig}, top right panel). We also find no correlation between $\xi$ and the outflow velocity $v_{\rm out}$ of the wind.

Another possibility for the \NH-$R$ relation could be the effect of the viewing angle of the AGN. According to some theoretical wind models, from high to low inclination angles (i.e. from edge-on to face-on view), the wind \NH is expected to decrease (e.g. \citealt{Prog04}). However, the inclination angle $i$ of the objects in our sample, which are all Seyfert-1 AGN, are rather limited to a small range (10$\degr$ to 40$\degr$). Moreover, we do not find a correlation between \NH and $i$ (Fig. \ref{NH_rel_fig}, middle left panel). Therefore, the inclination angle is unlikely to be the cause of the observed lowering in the wind \NH as $R$ increases.

One of the mechanisms for the launch of ionised winds in AGN is thermal evaporation from the AGN dusty torus (e.g. \citealt{Krol01}). Moreover, such winds from the torus may be boosted by the IR radiation pressure on dust grains in the wind (e.g. \citealt{Doro11}). In order to find out if there is a relation between the ionised wind and the dusty torus in our AGN sample, we checked for correlations between the wind parameters and the emission from the dusty torus, which peaks in the mid-IR energy band. Similar to the radio-loudness parameter $R$ of the jet, the `mid-IR loudness' parameter of the torus is a measure of the relative strength of the emission from the torus compared to the emission from the accretion disk, which we define as the ratio of mid-IR to optical ($B$) flux density ($F_{\rm mid \mhyphen IR} / F_{\rm opt}$). We find no significant correlation between $F_{\rm mid \mhyphen IR} / F_{\rm opt}$ and the wind parameters, including \NH, which is shown in Fig. \ref{NH_rel_fig}, middle right panel. Recent mid-IR interferometry observations have shown that significant fraction of the mid-IR emission in some AGN originates along the polar directions (e.g. \citealt{Honi13,Tris14}). This is thought to be associated to dusty winds from the AGN torus. However, the lack of a correlation between \NH and the mid-IR loudness parameter, as well as the no correlation between \NH and the inclination angle $i$, indicate that the AGN torus is unlikely to play a key role in the observed \NH variation in our sample and hence the observed \NH-$R$ relation.

The strong accretion-powered radiation in AGN can drive outflowing winds, provided the gas is not over-ionised (e.g. \citealt{Prog04,Prog07}). For such radiatively-driven winds, the mass outflow rate would be dependent on the mass accretion rate $\dot{M}$ and hence the Eddington ratio $\lambda$ of the AGN (e.g. \citealt{Kuro09}). Interestingly, an inverse relation between the radio-loudness parameter $R$ and the Eddington ratio $\lambda$ has been previously observed in radio-loud AGN (e.g. \citealt{Ho02,Siko07}). This relation shows that the radio power of the jet decreases with increasing $\dot{M}$ and $\lambda$. However, this relation is evident over a wide range of $\lambda$ (spanning several orders of magnitude), significantly more than the sub-Eddington $\lambda$ range of our AGN sample (spanning 1.7 orders of magnitude) as shown in Table \ref{properties_table} and Fig. \ref{NH_rel_fig} (bottom right panel). Moreover, we do not find a statistically significant correlation between $R$ and $\lambda$ for our AGN sample. Similarly, there is no significant correlation between the wind \NH and $\lambda$, nor between \NH and the black hole mass $M_{\rm BH}$, as shown in Fig. \ref{NH_rel_fig} (bottom panels). This means $\lambda$ of our selected AGN is effectively sampling a bin of the wider $R$-$\lambda$ relation reported in the literature. Therefore, the above suggest that our observed \NH-$R$ relation is not a consequence of large $\dot{M}$ and $\lambda$ variations in our AGN sample.

Since the different scenarios discussed earlier cannot provide a feasible explanation for the observed \NH-$R$ relation, we suggest that the most plausible explanation is associated to the only mechanism that can launch both jets and winds, i.e. the magnetic fields. Magnetic pressure generated in the accretion disk can drive winds (e.g. \citealt{Ohsu09}), while magnetocentrifugal acceleration along the field lines can drive jets aided by the spin of the black hole (e.g. \citealt{Blan77,Blan82}). Thus, depending on the black hole spin and the magnetic field configuration, the magnetic power maybe distributed differently over the jet and wind modes of the outflow. We discuss the magnetic explanation for the wind-jet \NH-$R$ relation in Sect. \ref{magnetic_sect}.

\subsection{Wind-jet connection by the magnetic fields in radio-loud AGN}
\label{magnetic_sect}

In Sect. \ref{explain_sect} we deduced that the \NH-$R$ relation is most likely the manifestation of the magnetic-powered outflow mechanism in radio-loud AGN. Interestingly, in black hole binaries (BHBs) the wind properties have been observed to change with the spectral state \citep{Neil09,Pont12,Mill12}. Observations show that while the wind is present in the high/soft state, it disappears in the low/hard state. The magnetic outflow mechanism has been proposed as a feasible explanation for the inverse relation between the winds and jets in BHBs \citep{Neil09,Pont12,Mill12}. In this scenario, the wind and the jet are driven by different configurations of the same magnetic field. In other words, the same magnetically-driven outflow is either a wind in the high/soft state, or a jet in the low/hard state (e.g. \citealt{Neil09}).

The wind-jet inverse relation phenomenon in BHBs is analogous to the one discovered in radio-loud AGN by our investigation. As discussed by \citet{Mill12}, the magnetic field configuration may change from toroidal to poloidal in transition from high/soft disk-dominated state to low/hard jet-dominated state. Thus, the magnetic outflow mechanism can provide the required explanation for the observed \NH-$R$ relation in radio-loud AGN. This is also further supported by the study of \citet{King13}, where they report that black hole winds and jets are regulated across the mass scale with a common launching mechanism driving both outflows, i.e. magnetohydrodynamics for winds and magnetocentrifugal forces for jets.

An important consequence of the above scenario is the likely changing geometry of the accretion disk with increasing jet activity. Disk truncation has been discussed by \citet{Lohf13} as an explanation for the jet cycle seen in 3C~120. From a multi-epoch case-study of this AGN in the radio, optical/UV, and X-rays, \citet{Lohf13} find a complete disk at one epoch, and a truncated/refilling disk at another epoch. The first epoch is seen as a rising trend in the optical/UV and X-ray emission from the disk, while the second epoch is associated to a dip in the disk emission and an increase in the radio jet activity. The behaviour of the ionised wind in 3C~120 has not been investigated in \citet{Lohf13}.

%
\begin{figure}[!tbp]
\centering
\hspace{-0.07cm}\resizebox{1.01\hsize}{!}{\includegraphics[angle=0]{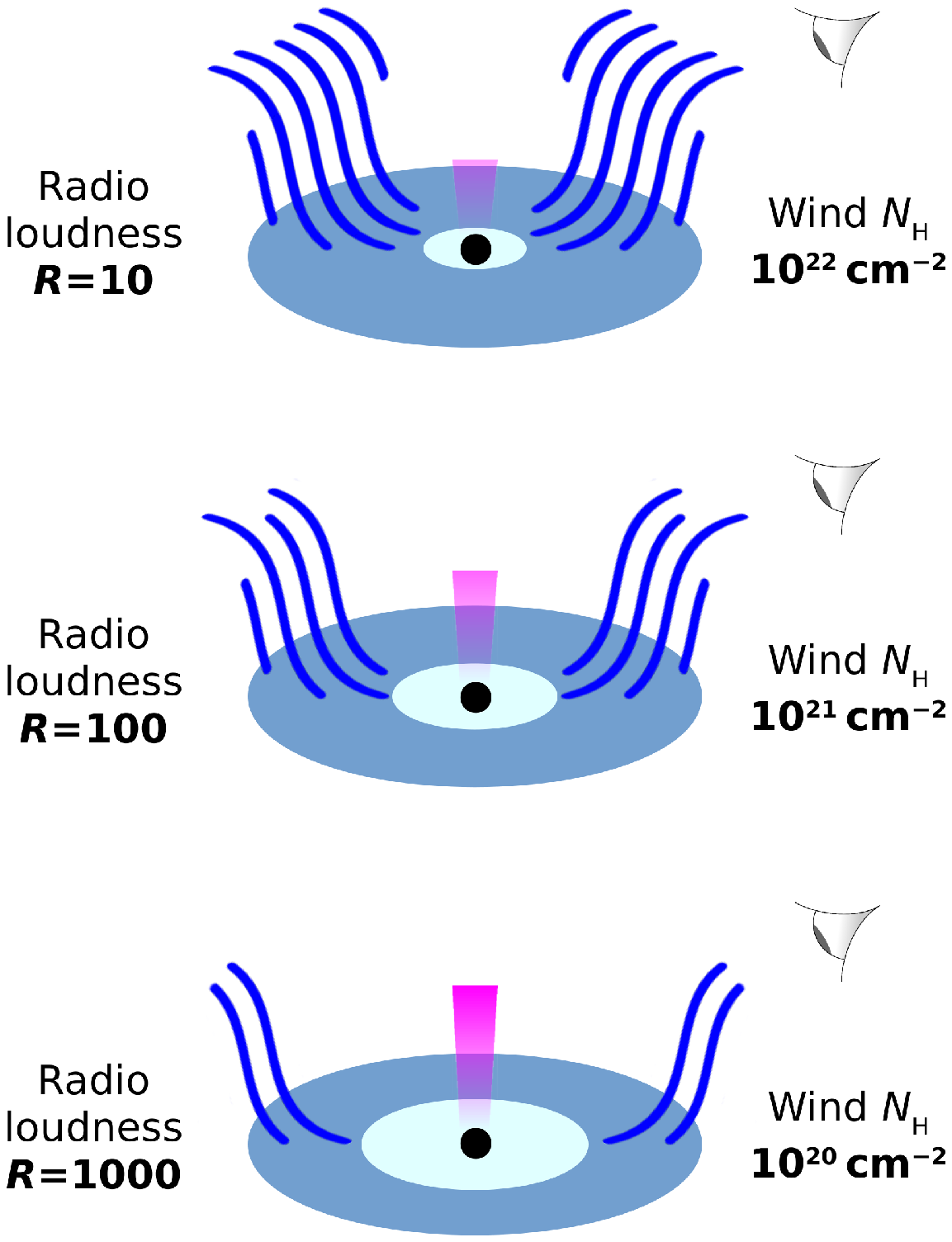}\hspace{-0.0cm}}
\caption{Illustration of the relation between the two modes of outflow (wind and jet) in radio-loud AGN, powered by different configurations of the magnetic outflow mechanism. The decrease in the observed column density \NH of the wind as the radio-loudness parameter $R$ of the jet increases is displayed.}
\label{cartoon}
\end{figure}

Finally, in Fig. \ref{cartoon} we illustrate the relation between the wind and the jet in radio-loud AGN based on the findings of our investigation. In radio-loud AGN with spinning black holes, stronger poloidal magnetic field and disk truncation give rise to brighter radio jets (i.e. higher $R$), and hence less toroidal magnetic winds are launched from the accretion disk (i.e. lower \NH).

In this study we considered only radio-loud AGN for investigating the wind-jet connection. However, in radio-quiet AGN, where winds are prominently present, there is also radio emission, albeit much fainter than in radio-loud AGN. At the low threshold limit of radio-loudness parameter ($R \approx 10$), we find that the wind \NH reaches about $10^{22}$~\cm (Fig. \ref{NH_R_fig}), which is a typical \NH seen in radio-quiet Seyfert-1 AGN (see e.g. \citealt{Blu05}). In radio-quiet AGN there is also a wide range of wind \NH (e.g. \citealt{Laha14}), which its association to the jet activity is yet to be established. This is however challenging because the nature and origin of radio emission in radio-quiet AGN is rather uncertain. There are different possible explanations given in the literature, such as coronal disk emission, star formation, emission from outflows including low-power jets, or perhaps a combination of them (see e.g. \citealt{Pane19} and references therein). Therefore, while ionised winds are commonly seen in radio-quiet AGN, investigating their link to the jet activity is complex since multiple processes may contribute to the faint radio emission in radio-quiet AGN.

The X-ray spectroscopy of ionised winds in radio-loud AGN is currently feasible for a limited number of X-ray bright sources as presented in this study (Fig. \ref{histo_fig}). The upcoming \athena X-ray observatory \citep{Nand13}, with its unprecedented X-ray sensitivity and energy resolution, enables us to extend the wind-jet study to a larger population of AGN that are not bright enough for the current X-ray observatories. It will also help us to simultaneously constrain all components of the wind using high-resolution spectroscopy in the soft and hard X-ray bands. The increase in the sample size will provide a more general characterisation of the wind-jet-disk connection in radio-loud AGN. 

\section{Conclusions}
\label{conclusions}

In this paper we have parameterised the ionised winds in a sample of 16 radio-loud AGN by carrying out high-resolution X-ray spectroscopy and photoionisation modelling. The results of our study shine new light on the relation between the two modes of outflow (wind and jet) in AGN. From the findings of our investigation we conclude the following. 

\begin{enumerate}

\item We discover that in radio-loud AGN there is a significant inverse correlation between the column density \NH of the ionised wind and the radio-loudness parameter $R$ of the jet. The wind \NH, ranging over ${\sim 10^{20}}$--${10^{22}}$~\cm, is seen to decrease as a power-law function with $R$, ranging over ${\sim 10}$--${10^{3}}$. There is no significant correlation between $R$ and the other wind parameters, namely the ionisation parameter $\xi$ and the outflow velocity $v_{\rm out}$.  
\medskip

\item The results of our study show that the observed \NH-$R$ relation is not caused by inclination or ionisation effects. There is also no association to the black hole mass and the Eddington luminosity ratio of the AGN. The \NH-$R$ relation is also independent of the AGN dusty torus.
\medskip

\item The most feasible explanation for the origin of the observed \NH-$R$ relation is the magnetic outflow mechanism in radio-loud AGN. Depending on the black hole spin and the magnetic field configuration (toroidal or poloidal), the magnetic power drives the wind or the jet mode of the outflow differently. Disk truncation is also a likely consequence of this scenario as the jet becomes stronger and the wind becomes weaker.
\medskip

\item The observed wind-jet bimodality in radio-loud AGN is analogous to that previously found in stellar-mass black holes. Our results point towards the magnetohydrodynamic mechanism for the ejection of ionised winds from the accretion disk in AGN.
\medskip

\end{enumerate}
\begin{acknowledgements}
M.M. and E.C. are supported by the Netherlands Organisation for Scientific Research (NWO) through the Innovational Research Incentives Scheme Vidi grant 639.042.525. This work is based on observations obtained with XMM-Newton, an ESA science mission with instruments and contributions directly funded by ESA Member States and the USA (NASA). This research made use of the cross-match service provided by CDS, Strasbourg. We thank R. Morganti for the useful discussions. We thank the anonymous referee for the constructive comments.
\end{acknowledgements}


\begin{thebibliography}{}

\bibitem[{{Ballantyne}(2005)}]{Ball05b}
{Ballantyne}, D.~R. 2005, \mnras, 362, 1183

\bibitem[{{Ballantyne} \& {Fabian}(2005)}]{Ball05a}
{Ballantyne}, D.~R. \& {Fabian}, A.~C. 2005, \apjl, 622, L97

\bibitem[{{Begelman} {et~al.}(1983){Begelman}, {McKee}, \& {Shields}}]{Bege83}
{Begelman}, M.~C., {McKee}, C.~F., \& {Shields}, G.~A. 1983, \apj, 271, 70

\bibitem[{{Behar} {et~al.}(2017){Behar}, {Peretz}, {Kriss}, {Kaastra}, {Arav},
  {Bianchi}, {Branduardi-Raymont}, {Cappi}, {Costantini}, {De Marco}, {Di
  Gesu}, {Ebrero}, {Kaspi}, {Mehdipour}, {Paltani}, {Petrucci}, {Ponti}, \&
  {Ursini}}]{Beh17}
{Behar}, E., {Peretz}, U., {Kriss}, G.~A., {et~al.} 2017, \aap, 601, A17

\bibitem[{{Bicay} {et~al.}(1995){Bicay}, {Kojoian}, {Seal}, {Dickinson}, \&
  {Malkan}}]{Bica95}
{Bicay}, M.~D., {Kojoian}, G., {Seal}, J., {Dickinson}, D.~F., \& {Malkan},
  M.~A. 1995, The Astrophysical Journal Supplement Series, 98, 369

\bibitem[{{Blandford} \& {Znajek}(1977)}]{Blan77}
{Blandford}, R.~D. \& {Znajek}, R.~L. 1977, \mnras, 179, 433

\bibitem[{{Blandford} \& {Payne}(1982)}]{Blan82}
{Blandford}, R.~D. \& {Payne}, D.~G. 1982, \mnras, 199, 883

\bibitem[{{Blustin} {et~al.}(2005){Blustin}, {Page}, {Fuerst},
  {Branduardi-Raymont}, \& {Ashton}}]{Blu05}
{Blustin}, A.~J., {Page}, M.~J., {Fuerst}, S.~V., {Branduardi-Raymont}, G., \&
  {Ashton}, C.~E. 2005, \aap, 431, 111

\bibitem[{{de Plaa} {et~al.}(2004){de Plaa}, {Kaastra}, {Tamura},
  {Pointecouteau}, {Mendez}, \& {Peterson}}]{dePl04}
{de Plaa}, J., {Kaastra}, J.~S., {Tamura}, T., {et~al.} 2004, \aap, 423, 49

\bibitem[{{den Herder} {et~al.}(2001){den Herder}, {Brinkman}, {Kahn},
  {Branduardi-Raymont}, {Thomsen}, , {et~al.}}]{denH01}
{den Herder}, J.~W., {Brinkman}, A.~C., {Kahn}, S.~M., {et~al.} 2001, \aap,
  365, L7

\bibitem[{{Di Gesu} \& {Costantini}(2016)}]{DiGe16}
{Di Gesu}, L. \& {Costantini}, E. 2016, \aap, 594, A88

\bibitem[{{Dodson} {et~al.}(2008){Dodson}, {Fomalont}, {Wiik}, {Horiuchi},
  {Hirabayashi}, {Edwards}, {Murata}, {Asaki}, {Moellenbrock}, {Scott},
  {Taylor}, {Gurvits}, {Paragi}, {Frey}, {Shen}, {Lovell}, {Tingay}, {Rioja},
  {Fodor}, {Lister}, {Mosoni}, {Coldwell}, {Piner}, \& {Yang}}]{Dods08}
{Dodson}, R., {Fomalont}, E.~B., {Wiik}, K., {et~al.} 2008, \apjs, 175, 314

\bibitem[{{Dorodnitsyn} {et~al.}(2011){Dorodnitsyn}, {Bisnovatyi-Kogan}, \&
  {Kallman}}]{Doro11}
{Dorodnitsyn}, A., {Bisnovatyi-Kogan}, G.~S., \& {Kallman}, T. 2011, \apj, 741,
  29

\bibitem[{{Doroshenko} {et~al.}(2012){Doroshenko}, {Sergeev}, {Klimanov},
  {Pronik}, \& {Efimov}}]{Doro12}
{Doroshenko}, V.~T., {Sergeev}, S.~G., {Klimanov}, S.~A., {Pronik}, V.~I., \&
  {Efimov}, Y.~S. 2012, \mnras, 426, 416

\bibitem[{{Fabian}(2012)}]{Fabi12}
{Fabian}, A.~C. 2012, \araa, 50, 455

\bibitem[{{Ferrarese} \& {Merritt}(2000)}]{Ferr00}
{Ferrarese}, L. \& {Merritt}, D. 2000, \apjl, 539, L9

\bibitem[{{Foschini} {et~al.}(2015){Foschini}, {Berton}, {Caccianiga}, {Ciroi},
  {Cracco}, {Peterson}, {Angelakis}, {Braito}, {Fuhrmann}, {Gallo}, {Grupe},
  {J{\"a}rvel{\"a}}, {Kaufmann}, {Komossa}, {Kovalev}, {L{\"a}hteenm{\"a}ki},
  {Lisakov}, {Lister}, {Mathur}, {Richards}, {Romano}, {Sievers},
  {Tagliaferri}, {Tammi}, {Tibolla}, {Tornikoski}, {Vercellone}, {La Mura},
  {Maraschi}, \& {Rafanelli}}]{Fosc15}
{Foschini}, L., {Berton}, M., {Caccianiga}, A., {et~al.} 2015, \aap, 575, A13

\bibitem[{{Fukumura} {et~al.}(2010){Fukumura}, {Kazanas}, {Contopoulos}, \&
  {Behar}}]{Fuku10}
{Fukumura}, K., {Kazanas}, D., {Contopoulos}, I., \& {Behar}, E. 2010, \apj,
  715, 636

\bibitem[{{Gear} {et~al.}(1994){Gear}, {Stevens}, {Hughes}, {Litchfield},
  {Robson}, {Terasranta}, {Valtaoja}, {Steppe}, {Aller}, \& {Aller}}]{Gear94}
{Gear}, W.~K., {Stevens}, J.~A., {Hughes}, D.~H., {et~al.} 1994, \mnras, 267,
  167

\bibitem[{{Giovannini} {et~al.}(2001){Giovannini}, {Cotton}, {Feretti}, {Lara},
  \& {Venturi}}]{Giov01}
{Giovannini}, G., {Cotton}, W.~D., {Feretti}, L., {Lara}, L., \& {Venturi}, T.
  2001, \apj, 552, 508

\bibitem[{{Grier} {et~al.}(2017){Grier}, {Pancoast}, {Barth}, {Fausnaugh},
  {Brewer}, {Treu}, \& {Peterson}}]{Grie17}
{Grier}, C.~J., {Pancoast}, A., {Barth}, A.~J., {et~al.} 2017, \apj, 849, 146

\bibitem[{{Higginbottom} {et~al.}(2014){Higginbottom}, {Proga}, {Knigge},
  {Long}, {Matthews}, \& {Sim}}]{Higg14}
{Higginbottom}, N., {Proga}, D., {Knigge}, C., {et~al.} 2014, \apj, 789, 19

\bibitem[{{Ho}(2002)}]{Ho02}
{Ho}, L.~C. 2002, \apj, 564, 120

\bibitem[{{Hogan} {et~al.}(2011){Hogan}, {Lister}, {Kharb}, {Marshall}, \&
  {Cooper}}]{Hoga11}
{Hogan}, B.~S., {Lister}, M.~L., {Kharb}, P., {Marshall}, H.~L., \& {Cooper},
  N.~J. 2011, \apj, 730, 92

\bibitem[{{H{\"o}nig} {et~al.}(2013){H{\"o}nig}, {Kishimoto}, {Tristram},
  {Prieto}, {Gandhi}, {Asmus}, {Antonucci}, {Burtscher}, {Duschl}, \&
  {Weigelt}}]{Honi13}
{H{\"o}nig}, S.~F., {Kishimoto}, M., {Tristram}, K.~R.~W., {et~al.} 2013, \apj,
  771, 87

\bibitem[{{Jansen} {et~al.}(2001){Jansen}, {Lumb}, {Altieri}, {Clavel}, {Ehle},
  , {et~al.}}]{Jans01}
{Jansen}, F., {Lumb}, D., {Altieri}, B., {et~al.} 2001, \aap, 365, L1

\bibitem[{{Jorstad} {et~al.}(2005){Jorstad}, {Marscher}, {Lister}, {Stirling},
  {Cawthorne}, {Gear}, {G{\'o}mez}, {Stevens}, {Smith}, {Forster}, \&
  {Robson}}]{Jors05}
{Jorstad}, S.~G., {Marscher}, A.~P., {Lister}, M.~L., {et~al.} 2005, \aj, 130,
  1418

\bibitem[{{Kaastra} \& {Barr}(1989)}]{Kaa89}
{Kaastra}, J.~S. \& {Barr}, P. 1989, \aap, 226, 59

\bibitem[{{Kaastra} {et~al.}(1996){Kaastra}, {Mewe}, \&
  {Nieuwenhuijzen}}]{Kaa96}
{Kaastra}, J.~S., {Mewe}, R., \& {Nieuwenhuijzen}, H. 1996, in UV and X-ray
  Spectroscopy of Astrophysical and Laboratory Plasmas, ed. K.~{Yamashita} \&
  T.~{Watanabe}, 411--414

\bibitem[{{Kaastra} {et~al.}(2000){Kaastra}, {Mewe}, {Liedahl}, {Komossa}, \&
  {Brinkman}}]{Kaa00}
{Kaastra}, J.~S., {Mewe}, R., {Liedahl}, D.~A., {Komossa}, S., \& {Brinkman},
  A.~C. 2000, \aap, 354, L83

\bibitem[{{Kaastra} {et~al.}(2012){Kaastra}, {Detmers}, {Mehdipour}, {Arav},
  {Behar}, {Bianchi}, {Branduardi-Raymont}, {Cappi}, {Costantini}, {Ebrero},
  {Kriss}, {Paltani}, {Petrucci}, {Pinto}, {Ponti}, {Steenbrugge}, \& {de
  Vries}}]{Kaas12}
{Kaastra}, J.~S., {Detmers}, R.~G., {Mehdipour}, M., {et~al.} 2012, \aap, 539,
  A117

\bibitem[{{Kaastra} \& {Bleeker}(2016)}]{Kaas16}
{Kaastra}, J.~S. \& {Bleeker}, J.~A.~M. 2016, \aap, 587, A151

\bibitem[{Kaastra {et~al.}(2017)Kaastra, Raassen, de~Plaa, \& Gu}]{Kaas17}
Kaastra, J.~S., Raassen, A. J.~J., de~Plaa, J., \& Gu, L. 2017, SPEX X-ray
  spectral fitting package, https://doi.org/10.5281/zenodo.2272992

\bibitem[{{Karamanavis}(2015)}]{Karam15}
{Karamanavis}, V.~. 2015, PhD thesis, Max-Planck-Institut f{\"u}r
  Radioastronomie

\bibitem[{{Kellermann} {et~al.}(1989){Kellermann}, {Sramek}, {Schmidt},
  {Shaffer}, \& {Green}}]{Kell89}
{Kellermann}, K.~I., {Sramek}, R., {Schmidt}, M., {Shaffer}, D.~B., \& {Green},
  R. 1989, \aj, 98, 1195

\bibitem[{{Kharb} {et~al.}(2006){Kharb}, {O'Dea}, {Baum}, {Colbert}, \&
  {Xu}}]{Khar06}
{Kharb}, P., {O'Dea}, C.~P., {Baum}, S.~A., {Colbert}, E.~J.~M., \& {Xu}, C.
  2006, \apj, 652, 177

\bibitem[{{King} {et~al.}(2013){King}, {Miller}, {Raymond}, {Fabian},
  {Reynolds}, {G{\"u}ltekin}, {Cackett}, {Allen}, {Proga}, \&
  {Kallman}}]{King13}
{King}, A.~L., {Miller}, J.~M., {Raymond}, J., {et~al.} 2013, \apj, 762, 103

\bibitem[{{King} \& {Pounds}(2015)}]{King15}
{King}, A. \& {Pounds}, K. 2015, \araa, 53, 115

\bibitem[{{K\"{o}nigl} \& {Kartje}(1994)}]{Koni94}
{K\"{o}nigl}, A. \& {Kartje}, J.~F. 1994, \apj, 434, 446

\bibitem[{{Krolik} {et~al.}(1981){Krolik}, {McKee}, \& {Tarter}}]{Kro81}
{Krolik}, J.~H., {McKee}, C.~F., \& {Tarter}, C.~B. 1981, \apj, 249, 422

\bibitem[{{Krolik} \& {Kriss}(2001)}]{Krol01}
{Krolik}, J.~H. \& {Kriss}, G.~A. 2001, \apj, 561, 684

\bibitem[{{Kuehr} {et~al.}(1981){Kuehr}, {Pauliny-Toth}, {Witzel}, \&
  {Schmidt}}]{Kueh81}
{Kuehr}, H., {Pauliny-Toth}, I.~I.~K., {Witzel}, A., \& {Schmidt}, J. 1981,
  \aj, 86, 854

\bibitem[{{Kurosawa} \& {Proga}(2009)}]{Kuro09}
{Kurosawa}, R. \& {Proga}, D. 2009, \mnras, 397, 1791

\bibitem[{{Laha} {et~al.}(2014){Laha}, {Guainazzi}, {Dewangan}, {Chakravorty},
  \& {Kembhavi}}]{Laha14}
{Laha}, S., {Guainazzi}, M., {Dewangan}, G.~C., {Chakravorty}, S., \&
  {Kembhavi}, A.~K. 2014, \mnras, 441, 2613

\bibitem[{{Laurent-Muehleisen} {et~al.}(1997){Laurent-Muehleisen}, {Kollgaard},
  {Ryan}, {Feigelson}, {Brinkmann}, \& {Siebert}}]{Laur97}
{Laurent-Muehleisen}, S.~A., {Kollgaard}, R.~I., {Ryan}, P.~J., {et~al.} 1997,
  \aaps, 122, 235

\bibitem[{{Lodders} {et~al.}(2009){Lodders}, {Palme}, \& {Gail}}]{Lod09}
{Lodders}, K., {Palme}, H., \& {Gail}, H.-P. 2009, Landolt B{\"o}rnstein, 44

\bibitem[{{Lohfink} {et~al.}(2013){Lohfink}, {Reynolds}, {Jorstad}, {Marscher},
  {Miller}, {Aller}, {Aller}, {Brenneman}, {Fabian}, {Miller}, {Mushotzky},
  {Nowak}, \& {Tombesi}}]{Lohf13}
{Lohfink}, A.~M., {Reynolds}, C.~S., {Jorstad}, S.~G., {et~al.} 2013, \apj,
  772, 83

\bibitem[{{Magdziarz} \& {Zdziarski}(1995)}]{Magd95}
{Magdziarz}, P. \& {Zdziarski}, A.~A. 1995, \mnras, 273, 837

\bibitem[{{Marchesini} {et~al.}(2004){Marchesini}, {Celotti}, \&
  {Ferrarese}}]{Marc04}
{Marchesini}, D., {Celotti}, A., \& {Ferrarese}, L. 2004, \mnras, 351, 733

\bibitem[{{Mehdipour} {et~al.}(2011){Mehdipour}, {Branduardi-Raymont},
  {Kaastra}, {Petrucci}, {Kriss}, , {et~al.}}]{Meh11}
{Mehdipour}, M., {Branduardi-Raymont}, G., {Kaastra}, J.~S., {et~al.} 2011,
  \aap, 534, A39

\bibitem[{{Mehdipour} {et~al.}(2015){Mehdipour}, {Kaastra}, {Kriss}, {Cappi},
  {Petrucci}, {Steenbrugge}, {Arav}, {Behar}, {Bianchi}, {Boissay},
  {Branduardi-Raymont}, {Costantini}, {Ebrero}, {Di Gesu}, {Harrison}, {Kaspi},
  {De Marco}, {Matt}, {Paltani}, {Peterson}, {Ponti}, {Pozo Nu{\~n}ez}, {De
  Rosa}, {Ursini}, {de Vries}, {Walton}, \& {Whewell}}]{Meh15a}
{Mehdipour}, M., {Kaastra}, J.~S., {Kriss}, G.~A., {et~al.} 2015, \aap, 575,
  A22

\bibitem[{{Mehdipour} {et~al.}(2016){Mehdipour}, {Kaastra}, \&
  {Kallman}}]{Meh16b}
{Mehdipour}, M., {Kaastra}, J.~S., \& {Kallman}, T. 2016, \aap, 596, A65

\bibitem[{{Mehdipour} \& {Costantini}(2018)}]{Meh18b}
{Mehdipour}, M. \& {Costantini}, E. 2018, \aap, 619, A20

\bibitem[{{Miller} {et~al.}(2012){Miller}, {Raymond}, {Fabian}, {Reynolds},
  {King}, {Kallman}, {Cackett}, {van der Klis}, \& {Steeghs}}]{Mill12}
{Miller}, J.~M., {Raymond}, J., {Fabian}, A.~C., {et~al.} 2012, \apjl, 759, L6

\bibitem[{{Mingo} {et~al.}(2014){Mingo}, {Hardcastle}, {Croston}, {Dicken},
  {Evans}, {Morganti}, \& {Tadhunter}}]{Ming14}
{Mingo}, B., {Hardcastle}, M.~J., {Croston}, J.~H., {et~al.} 2014, \mnras, 440,
  269

\bibitem[{{Mitsuda} {et~al.}(2007){Mitsuda}, {Bautz}, {Inoue}, {Kelley},
  {Koyama}, {Kunieda}, {Makishima}, {Ogawara}, {Petre}, {Takahashi}, {Tsunemi},
  {White}, {Anabuki}, {Angelini}, {Arnaud}, {Awaki}, {Bamba}, {Boyce}, {Brown},
  {Chan}, {Cottam}, {Dotani}, {Doty}, {Ebisawa}, {Ezoe}, {Fabian}, {Figueroa},
  {Fujimoto}, {Fukazawa}, {Furusho}, {Furuzawa}, {Gendreau}, {Griffiths},
  {Haba}, {Hamaguchi}, {Harrus}, {Hasinger}, {Hatsukade}, {Hayashida}, {Henry},
  {Hiraga}, {Holt}, {Hornschemeier}, {Hughes}, {Hwang}, {Ishida}, {Ishisaki},
  {Isobe}, {Itoh}, {Iyomoto}, {Kahn}, {Kamae}, {Katagiri}, {Kataoka},
  {Katayama}, {Kawai}, {Kilbourne}, {Kinugasa}, {Kissel}, {Kitamoto}, {Kohama},
  {Kohmura}, {Kokubun}, {Kotani}, {Kotoku}, {Kubota}, {Madejski}, {Maeda},
  {Makino}, {Markowitz}, {Matsumoto}, {Matsumoto}, {Matsuoka}, {Matsushita},
  {McCammon}, {Mihara}, {Misaki}, {Miyata}, {Mizuno}, {Mori}, {Mori}, {Morii},
  {Moseley}, {Mukai}, {Murakami}, {Murakami}, {Mushotzky}, {Nagase}, {Namiki},
  {Negoro}, {Nakazawa}, {Nousek}, {Okajima}, {Ogasaka}, {Ohashi}, {Oshima},
  {Ota}, {Ozaki}, {Ozawa}, {Parmar}, {Pence}, {Porter}, {Reeves}, {Ricker},
  {Sakurai}, {Sanders}, {Senda}, {Serlemitsos}, {Shibata}, {Soong}, {Smith},
  {Suzuki}, {Szymkowiak}, {Takahashi}, {Tamagawa}, {Tamura}, {Tamura},
  {Tanaka}, {Tashiro}, {Tawara}, {Terada}, {Terashima}, {Tomida}, {Torii},
  {Tsuboi}, {Tsujimoto}, {Tsuru}, {Turner}, {Ueda}, {Ueno}, {Ueno}, {Uno},
  {Urata}, {Watanabe}, {Yamamoto}, {Yamaoka}, {Yamasaki}, {Yamashita},
  {Yamauchi}, {Yamauchi}, {Yaqoob}, {Yonetoku}, \& {Yoshida}}]{Mits07}
{Mitsuda}, K., {Bautz}, M., {Inoue}, H., {et~al.} 2007, \pasj, 59, S1

\bibitem[{{Morganti}(2017)}]{Morg17}
{Morganti}, R. 2017, Frontiers in Astronomy and Space Sciences, 4, 42

\bibitem[{{Moshir} {et~al.}(1990){Moshir}, {Copan}, {Conrow}, {McCallon},
  {Hacking}, {Gregorich}, {Rohrbach}, {Melnyk}, {Rice}, \& {Fullmer}}]{Mosh90}
{Moshir}, M., {Copan}, G., {Conrow}, T., {et~al.} 1990, in IRAS Faint Source
  Catalogue, version 2.0 (1990)

\bibitem[{{Nandra} {et~al.}(2013){Nandra}, {Barret}, {Barcons}, {Fabian}, {den
  Herder}, {Piro}, {Watson}, {Adami}, {Aird}, {Afonso}, \& et~al.}]{Nand13}
{Nandra}, K., {Barret}, D., {Barcons}, X., {et~al.} 2013, ArXiv e-prints
  [\eprint[arXiv]{1306.2307}]

\bibitem[{{Neilsen} \& {Lee}(2009)}]{Neil09}
{Neilsen}, J. \& {Lee}, J.~C. 2009, \nat, 458, 481

\bibitem[{{Ohsuga} {et~al.}(2009){Ohsuga}, {Mineshige}, {Mori}, \&
  {Kato}}]{Ohsu09}
{Ohsuga}, K., {Mineshige}, S., {Mori}, M., \& {Kato}, Y. 2009, \pasj, 61, L7

\bibitem[{{Page} {et~al.}(2003){Page}, {O'Brien}, {Reeves}, \&
  {Breeveld}}]{Page03}
{Page}, K.~L., {O'Brien}, P.~T., {Reeves}, J.~N., \& {Breeveld}, A.~A. 2003,
  \mnras, 340, 1052

\bibitem[{{Paltani} \& {T{\"u}rler}(2005)}]{Palt05}
{Paltani}, S. \& {T{\"u}rler}, M. 2005, \aap, 435, 811

\bibitem[{{Panessa} {et~al.}(2019){Panessa}, {Baldi}, {Laor}, {Padovani},
  {Behar}, \& {McHardy}}]{Pane19}
{Panessa}, F., {Baldi}, R.~D., {Laor}, A., {et~al.} 2019, arXiv e-prints
  [\eprint[arXiv]{1902.05917}]

\bibitem[{{Peterson} {et~al.}(2004){Peterson}, {Ferrarese}, {Gilbert}, {Kaspi},
  {Malkan}, , {et~al.}}]{Pet04}
{Peterson}, B.~M., {Ferrarese}, L., {Gilbert}, K.~M., {et~al.} 2004, \apj, 613,
  682

\bibitem[{{Pineau} {et~al.}(2011){Pineau}, {Boch}, \& {Derriere}}]{Pine11}
{Pineau}, F.-X., {Boch}, T., \& {Derriere}, S. 2011, in Astronomical Society of
  the Pacific Conference Series, Vol. 442, Astronomical Data Analysis Software
  and Systems XX, ed. I.~N. {Evans}, A.~{Accomazzi}, D.~J. {Mink}, \& A.~H.
  {Rots}, 85

\bibitem[{{Piotrovich} {et~al.}(2015){Piotrovich}, {Gnedin}, {Silant'ev},
  {Natsvlishvili}, \& {Buliga}}]{Piot15}
{Piotrovich}, M.~Y., {Gnedin}, Y.~N., {Silant'ev}, N.~A., {Natsvlishvili},
  T.~M., \& {Buliga}, S.~D. 2015, \mnras, 454, 1157

\bibitem[{{Ponti} {et~al.}(2012){Ponti}, {Fender}, {Begelman}, {Dunn},
  {Neilsen}, \& {Coriat}}]{Pont12}
{Ponti}, G., {Fender}, R.~P., {Begelman}, M.~C., {et~al.} 2012, \mnras, 422,
  L11

\bibitem[{{Proga} \& {Kallman}(2004)}]{Prog04}
{Proga}, D. \& {Kallman}, T.~R. 2004, \apj, 616, 688

\bibitem[{{Proga}(2007)}]{Prog07}
{Proga}, D. 2007, \apj, 661, 693

\bibitem[{{Reeves} {et~al.}(2009){Reeves}, {Sambruna}, {Braito}, \&
  {Eracleous}}]{Reev09b}
{Reeves}, J.~N., {Sambruna}, R.~M., {Braito}, V., \& {Eracleous}, M. 2009,
  \apjl, 702, L187

\bibitem[{{Reynolds}(1997)}]{Reyn97}
{Reynolds}, C.~S. 1997, \mnras, 286, 513

\bibitem[{{Rokaki} {et~al.}(2003){Rokaki}, {Lawrence}, {Economou}, \&
  {Mastichiadis}}]{Roka03}
{Rokaki}, E., {Lawrence}, A., {Economou}, F., \& {Mastichiadis}, A. 2003,
  \mnras, 340, 1298

\bibitem[{{Schmidt} \& {Green}(1983)}]{Schm83}
{Schmidt}, M. \& {Green}, R.~F. 1983, \apj, 269, 352

\bibitem[{{Shimmins} \& {Bolton}(1972)}]{Shim72}
{Shimmins}, A.~J. \& {Bolton}, J.~G. 1972, Australian Journal of Physics
  Astrophysical Supplement, 23, 1

\bibitem[{{Sikora} {et~al.}(2007){Sikora}, {Stawarz}, \& {Lasota}}]{Siko07}
{Sikora}, M., {Stawarz}, {\L}., \& {Lasota}, J.-P. 2007, \apj, 658, 815

\bibitem[{{Stawarz}(2004)}]{Staw04}
{Stawarz}, {\L}. 2004, \apj, 613, 119

\bibitem[{{Steenbrugge} {et~al.}(2005){Steenbrugge}, {Kaastra}, {Crenshaw},
  {Kraemer}, {Arav}, {George}, {Liedahl}, {van der Meer}, {Paerels}, {Turner},
  \& {Yaqoob}}]{Stee05}
{Steenbrugge}, K.~C., {Kaastra}, J.~S., {Crenshaw}, D.~M., {et~al.} 2005, \aap,
  434, 569

\bibitem[{{Str{\"u}der} {et~al.}(2001){Str{\"u}der}, {Briel}, {Dennerl},
  {Hartmann}, {Kendziorra}, , {et~al.}}]{Stru01}
{Str{\"u}der}, L., {Briel}, U., {Dennerl}, K., {et~al.} 2001, \aap, 365, L18

\bibitem[{{Tanaka} {et~al.}(1994){Tanaka}, {Inoue}, \& {Holt}}]{Tana94}
{Tanaka}, Y., {Inoue}, H., \& {Holt}, S.~S. 1994, \pasj, 46, L37

\bibitem[{{Tarter} {et~al.}(1969){Tarter}, {Tucker}, \& {Salpeter}}]{Tar69}
{Tarter}, C.~B., {Tucker}, W.~H., \& {Salpeter}, E.~E. 1969, \apj, 156, 943

\bibitem[{{Tombesi} {et~al.}(2014){Tombesi}, {Tazaki}, {Mushotzky}, {Ueda},
  {Cappi}, {Gofford}, {Reeves}, \& {Guainazzi}}]{Tomb14}
{Tombesi}, F., {Tazaki}, F., {Mushotzky}, R.~F., {et~al.} 2014, \mnras, 443,
  2154

\bibitem[{{Torresi} {et~al.}(2012){Torresi}, {Grandi}, {Costantini}, \&
  {Palumbo}}]{Torr12}
{Torresi}, E., {Grandi}, P., {Costantini}, E., \& {Palumbo}, G.~G.~C. 2012,
  \mnras, 419, 321

\bibitem[{{Tristram} {et~al.}(2014){Tristram}, {Burtscher}, {Jaffe},
  {Meisenheimer}, {H{\"o}nig}, {Kishimoto}, {Schartmann}, \&
  {Weigelt}}]{Tris14}
{Tristram}, K.~R.~W., {Burtscher}, L., {Jaffe}, W., {et~al.} 2014, \aap, 563,
  A82

\bibitem[{{Turner} {et~al.}(2001){Turner}, {Abbey}, {Arnaud}, {Balasini},
  {Barbera}, , {et~al.}}]{Turn01}
{Turner}, M.~J.~L., {Abbey}, A., {Arnaud}, M., {et~al.} 2001, \aap, 365, L27

\bibitem[{{Urry} \& {Padovani}(1995)}]{Urry95}
{Urry}, C.~M. \& {Padovani}, P. 1995, \pasp, 107, 803

\bibitem[{{Vasudevan} \& {Fabian}(2007)}]{Vasu07}
{Vasudevan}, R.~V. \& {Fabian}, A.~C. 2007, \mnras, 381, 1235

\bibitem[{{Veilleux} {et~al.}(2009){Veilleux}, {Kim}, {Rupke}, {Peng},
  {Tacconi}, {Genzel}, {Lutz}, {Sturm}, {Contursi}, {Schweitzer}, {Dasyra},
  {Ho}, {Sanders}, \& {Burkert}}]{Veil09}
{Veilleux}, S., {Kim}, D.-C., {Rupke}, D.~S.~N., {et~al.} 2009, \apj, 701, 587

\bibitem[{{V{\'e}ron-Cetty} \& {V{\'e}ron}(2010)}]{Vero10}
{V{\'e}ron-Cetty}, M.~P. \& {V{\'e}ron}, P. 2010, \aap, 518, A10

\bibitem[{{Willingale} {et~al.}(2013){Willingale}, {Starling}, {Beardmore},
  {Tanvir}, \& {O'Brien}}]{Will13}
{Willingale}, R., {Starling}, R.~L.~C., {Beardmore}, A.~P., {Tanvir}, N.~R., \&
  {O'Brien}, P.~T. 2013, \mnras, 431, 394

\bibitem[{{Woo} \& {Urry}(2002)}]{Woo02}
{Woo}, J.-H. \& {Urry}, C.~M. 2002, \apj, 579, 530

\bibitem[{{Worrall} {et~al.}(2012){Worrall}, {Birkinshaw}, {Young}, {Momtahan},
  {Fosbury}, {Morganti}, {Tadhunter}, \& {Verdoes Kleijn}}]{Worr12}
{Worrall}, D.~M., {Birkinshaw}, M., {Young}, A.~J., {et~al.} 2012, \mnras, 424,
  1346

\bibitem[{{Wright} \& {Otrupcek}(1990)}]{Wrig90}
{Wright}, A. \& {Otrupcek}, R. 1990, PKS Catalog (1990), 0

\bibitem[{{Zycki} \& {Czerny}(1994)}]{Zyck94}
{Zycki}, P.~T. \& {Czerny}, B. 1994, \mnras, 266, 653

\bibitem[{{Zycki} {et~al.}(1999){Zycki}, {Done}, \& {Smith}}]{Zyck99}
{Zycki}, P.~T., {Done}, C., \& {Smith}, D.~A. 1999, \mnras, 305, 231

\end{thebibliography}
\end{document}